\documentclass[lettersize,journal]{IEEEtran}
\usepackage{amsmath,amssymb,amsfonts}
\usepackage{algorithmic}
\usepackage{array}
\usepackage[caption=false,font=normalsize,labelfont=sf,textfont=sf]{subfig}
\usepackage{textcomp}
\usepackage{stfloats}
\usepackage{url}
\usepackage{verbatim}
\usepackage{graphicx}
\usepackage{bm}
\usepackage{hyperref}
\usepackage{cases}
\usepackage[linesnumbered,ruled,vlined]{algorithm2e}

\graphicspath{{Figs/}}

\begin{document}

\title{Uncertainty-Aware Jamming Mitigation with\\Active RIS: A Robust Stackelberg Game Approach}

\author{Xiao Tang, Zhen Ma, Limeng Dong, Yichen Wang, Qinghe Du, Dusit Niyato, and Zhu Han%
\thanks{X. Tang is with the School of Information and Communication Engineering, Xi'an Jiaotong University, Xi'an 710049, China, and also with Shenzhen Research Institute of Northwestern Polytechnical University, Shenzhen 518057, China. (e-mail: tangxiao@xjtu.edu.cn)}% <-this % stops a space
\thanks{Z. Ma and L. Dong are with the School of Electronics and Information, Northwestern Polytechnical University, Xi'an 710072, China.}% <-this % stops a space
\thanks{Y. Wang and Q. Du are with the School of Information and Communication Engineering, Xi'an Jiaotong University, Xi'an 710049, China.}%
\thanks{D. Niyato is with the College of Computing and Data Science, Nanyang Technological University, Singapore.}%
\thanks{Z. Han is with the Department of Electrical and Computer Engineering, University of Houston, Houston 77004, USA, and also with the Department of Computer Science and Engineering, Kyung Hee University, Seoul 446-701, South Korea.}%
}

\maketitle

\begin{abstract}
Malicious jamming presents a pervasive threat to the secure communications, where the challenge becomes increasingly severe due to the growing capability of the jammer allowing the adaptation to legitimate transmissions. This paper investigates the jamming mitigation by leveraging an active reconfigurable intelligent surface (ARIS), where the channel uncertainties are particularly addressed for robust anti-jamming design. Towards this issue, we adopt the Stackelberg game formulation to model the strategic interaction between the legitimate side and the adversary, acting as the leader and follower, respectively. We prove the existence of the game equilibrium and adopt the backward induction method for equilibrium analysis. We first derive the optimal jamming policy as the follower's best response, which is then incorporated into the legitimate-side optimization for robust anti-jamming design. We address the uncertainty issue and reformulate the legitimate-side problem by exploiting the error bounds to combat the worst-case jamming attacks. The problem is decomposed within a block successive upper bound minimization (BSUM) framework to tackle the power allocation, transceiving beamforming, and active reflection, respectively, which are iterated towards the robust jamming mitigation scheme. Simulation results are provided to demonstrate the effectiveness of the proposed scheme in protecting the legitimate transmissions under uncertainties, and the superior performance in terms of jamming mitigation as compared with the baselines.
\end{abstract}


\section{Introduction} \label{sec:intro}

Security and reliability are fundamental to wireless communications, which are yet challenged by the jamming attacks given the open and shared nature of the wireless medium~\cite{back-jam}. The jamming attacks widely exist in wireless networks and can be conducted in various manners, presenting a pervasive threat to information security~\cite{back-jam-vis}. With the rapid development of wireless technologies and devices, the jamming equipment is also evolving, e.g., enabled with multiple antennas and the capability of adaptation to legitimate transmissions, posing an increasingly concerning threat for wireless systems. In this regard, the malicious jammer can also exploit spatial diversity, beamforming gains, and power adaptation to launch more focused and efficient attacks, significantly degrading communication performance beyond conventional simple constant jamming~\cite{back-adp-jam}. Therefore, ensuring secure transmissions against these ever-evolving and advanced jamming attacks is a critical yet challenging task, driving intensive research in wireless communication security~\cite{back-anti-jam}.

Conventionally, the jamming mitigation is achieved through resources diversity or advanced signal computation~\cite{jam-survey}. Despite the effectiveness, the anti-jamming performance of these approaches may still be bottlenecked by the limited network resources and device capabilities~\cite{jam-spec}. In this regard, the reconfigurable intelligent surface (RIS), as an emerging technology towards future 6G communications, has paved a new way to combat jamming attacks. With arrays of reflecting elements for on-demand radio environment manipulation, the RIS reflection allows a new dimension to enhance the legitimate transmissions~\cite{jam-sec}. Moreover, with the evolution of RIS technology, the reflecting elements are enabled with active amplification, noted as an active RIS (ARIS), where the amplification capability overcomes the ``double-fading'' deficit suffered at the passive RISs. Accordingly, the ARIS allows joint reflection and amplification for more proactive and flexible wireless environment programming, and thus offers superior resilience performance~\cite{jam-aris}.

Meanwhile, with the growing capability of jammers, the jamming attacks are able to adapt to the legitimate transmission behaviors for more effective attacks, no longer the simple constant jamming model~\cite{game-reac}. In this regard, the legitimate transmissions and jamming attacks are mutually affected, which necessities the investigation of their interactive results, rather than the single-sided optimization~\cite{game-gam}. Accordingly, the game-based analysis becomes instrumental with structured model of the strategic interactions between the legitimate system and adversary~\cite{game-jam}. When further probing into the ARIS-assisted anti-jamming communications, the reflection and amplification need to be incorporated into the game framework, which induces more complicated interactions with the attacking and counter-attacking behaviors. Consequently, the game model becomes significantly more intricate with multi-party involvement and the anti-jamming strategy deserves thorough investigation and elaborate design~\cite{game-ris}.

Moreover, the effective jamming mitigation is challenged by the information uncertainty issue~\cite{rob-back}. Due to the conflicting interest between the legitimate side and the adversary, it is rather difficult to obtain accurate channel state information in the system, particularly for the links associated with the non-cooperative jammer, which further impedes anti-jamming decision making. When a RIS is additionally involved, the large array of reflecting elements induces a high-dimensional channel, which introduces a greater number of parameters to estimate and thus further aggravates this issue~\cite{rob-ris}. Consequently, it is essential to investigate jamming mitigation design while incorporating uncertainty treatment to guarantee robust anti-jamming performance against worst-case conditions~\cite{rob-rob}. This discrepancy also underscores the need to address the jamming interactions within a game framework, since the adversary is adaptive and the channel information is imperfect, a robust solution requires modeling the strategic decision-making on both sides in the presence of uncertainties~\cite{rob-todo}.

Motivated by the requirement for information security under jamming attacks through an ARIS, in this paper, we employ the Stackelberg game-based formulation to investigate the anti-jamming interactions while considering the channel uncertainties. Particularly, the main contributions of this paper are summarized as follows:
\begin{itemize}
	\item We consider the legitimate transmissions in the presence of a malicious jammer, where an ARIS is deployed for jamming mitigation. We employ the two-stage Stackelberg game formulation where the legitimate system first commits to its defense strategy regarding transmission and ARIS configuration, and then the jammer adapts its attack behavior after observing the transmissions.
	\item For the formulated game, we confirm the existence of the Stackelberg equilibrium, follow which we employ the backward induction method and derive the optimal jamming strategy. Meanwhile, we formulate the robust ARIS-assisted jamming mitigation by incorporating the jammer's best response and channel uncertainties into consideration. We tackle the uncertainties with their bound information and reformulate the robust optimization to combat the worst-case jamming attacks.
	\item We decompose the legitimate-side problem within a block successive upper bound minimization (BSUM) framework to address the power allocation, transceive beamforming, and active reflection optimization, respectively. The subproblems are solved with their tight surrogate counterparts through successive convex approximation (SCA), and the blocked updates are conducted iteratively to reach the robust anti-jamming solution.
	\item We provide comprehensive simulation results to evaluate the performance of our proposed robust ARIS-assisted anti-jamming scheme. The results demonstrate that our proposed approach achieves superior jamming mitigation under various conditions as compared with the baselines, highlighting its resilience and effectiveness in the presence of channel uncertainties.
\end{itemize}

The remainder of this paper is organized as follows. Sec.~\ref{sec:rw} reviews the related work in the literature. Sec.~\ref{sec:sys} describes the anti-jamming communication system with uncertainty model. Sec.~\ref{sec:game} present the robust Stackelberg formulation along the equilibrium analysis. Sec.~\ref{sec:opt} details the proposed robust ARIS-assisted anti-jamming strategy. Sec.~\ref{sec:sim} provides simulation results, and finally, Sec.~\ref{sec:con} concludes this paper.

\section{Related Work} \label{sec:rw}

\subsection{RIS-Assisted Jamming Mitigation}

The anti-jamming communication keeps evolving alongside advanced wireless techniques~\cite{back-jam,s-suv}, with various strategies designed based on sophisticated optimization~\cite{ris-opt} or learning~\cite{ris-ln}. While effective to some certain degrees, these approaches are often limited by the available network resources and hardware capabilities~\cite{ris-back-jam1}. In this regard, the advent of RIS technology has introduced a new paradigm against jamming attacks. Accordingly, the existing studies have exploited RISs for anti-jamming designs with fairness guarantee~\cite{ris-fair}, wirelessly powered transmissions~\cite{ris-wpt}, and integrated communication and sensing~\cite{ris-isac}. However, the performance of these passive RIS-assisted proposals is fundamentally limited by the "double-fading" effect, which is likely to induce significant power loss in reflections~\cite{ris-a}. The concept of ARIS is proposed to amplify the reflected signal for jamming mitigation. Thus, recent work resorts to ARIS for self-sustainable anti-jamming protections~\cite{ris-sus}, proposes the joint use of active and passive RIS for security~\cite{ris-cas}, and adopts the multi-functional RIS to defend against jamming attacks~\cite{ris-sem}.

\subsection{Jamming Games with Uncertainties}

Modeling the strategic conflict between a legitimate system and an adversary is crucial for effective countermeasures, especially for adaptive and intelligent jammers~\cite{game-reac,game-jam}. In this respect, game theory provides a powerful framework, for which the widely applied zero-sum game has captured the direct confrontation~\cite{ga-zero1}. As the jamming scenario becomes more complex and involved, the Stackelberg game with a hierarchical structure appears more attractive~\cite{ga-sta1,ga-sta2}. Recently, along with the application of RISs in anti-jamming communications, it is also considered within the game-based formulations, e.g., the slotted jamming mitigation with RIS assistance~\cite{game-ris}, the RIS selection and reflection within a jamming game~\cite{ga-ris1}, and the learning-aided jamming defense with RISs~\cite{ga-ris2}. Furthermore, the channel uncertainties also presents a significant challenge for jamming mitigation design, particularly in the game context~\cite{rob-back,rob-rob}. Accordingly, the channel uncertainty is investigated for the robust protection against the worst-case attacks, e.g., the robust Stackelberg equilibrium for beam-domain anti-jamming transmission~\cite{ga-rob1}, and the robust fictitious play-based spectrum access policy~\cite{ga-rob2}. Moreover, with RIS further magnifying the difficulties with channel information acquisition, the recent studies also address the RIS-involved robust designs, e.g., the learning-based adaptation for reflection adaptation with uncertainties~\cite{ga-rob-ris1}, and the robust RIS-assisted jamming mitigation strategy~\cite{ga-rob-ris2}.

In a nutshell, the RIS-aided jamming mitigation is a promising approach for wireless security. While anti-jamming interactions have been actively investigated, a unified treatment that simultaneously accounts for a reactive jammer in an ARIS communication scenarios with uncertainties remains under-explored. This motivates our study to bridge this gap by exploiting the robust Stackelberg game for ARIS-assisted anti-jamming communications in this work.

\section{System Model} \label{sec:sys}

\subsection{System Setup}
\begin{figure}[t] \vspace{8pt}
    \centering
    \includegraphics[width=0.75\linewidth]{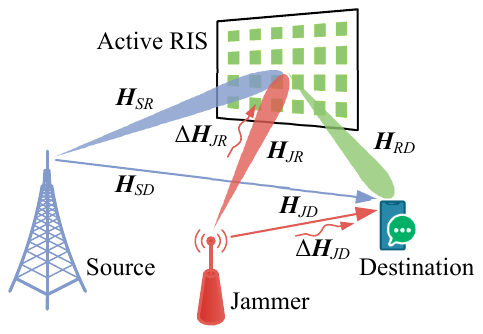} % Replace with your figure file
    \caption{System model of an ARIS-assisted communications against a jammer. The legitimate signals and jamming attacks reach the destination through the direct links (shown as arrows) and reflection-based links (shown as tapered ribbons).}
    \label{fig:sys}
\end{figure}

We consider a wireless communication system where a legitimate source node $S$ transmits information to a destination node $D$. This transmission is subject to intentional disruption by a malicious jammer $J$. To enhance the legitimate communication link and mitigate the jamming attacks, an ARIS with the capability of signal amplification and reflection, is deployed in the system, denoted by $R$. The system configuration is illustrated in Fig.~\ref{fig:sys}. The source node $S$, destination node $D$, and jammer $J$ are equipped with $N_S$, $N_D$, and $N_J$ antennas, respectively. The ARIS comprises $N$ reflecting elements, indexed by the set $\mathcal{N}=\{1,2,\ldots,N\}$. Each element of the ARIS can independently adjust the amplitude and phase of the incident signal. Basically, the considered system adopts an equivalent complex baseband MIMO model, and thus the formulation is not restricted to a specific communication standard or a fixed carrier frequency band. Let $\bm{H}_{SD} \in \mathbb{C}^{N_D \times N_S}$, $\bm{H}_{SR} \in \mathbb{C}^{N \times N_S}$, $\bm{H}_{RD} \in \mathbb{C}^{N_D \times N}$, $\bm{H}_{JD} \in \mathbb{C}^{N_D \times N_J}$, and $\bm{H}_{JR} \in \mathbb{C}^{N \times N_J}$ denote the channel matrices for the $S \to D$, $S \to R$, $R \to D$, $J \to D$, and $J \to R$ links, respectively. The ARIS employs a diagonal reflection matrix $\bm{\Theta} = \mathsf{diag}(\bm{\theta})$ with $\bm{\theta} = [\lambda_n e^{j\vartheta_n}]_{n\in\mathcal{N}}^T$, where $\lambda_n$ and $\vartheta_n$ are the reflection amplitude and phase shift induced by the $n$-th RIS element, respectively, and satisfies $ \lambda_n \le \lambda^{\max} $ and $\vartheta_n \in [0,2\pi)$, $ \forall n\in\mathcal{N} $, with $\lambda^{\max}$ being the maximum allowed amplification factor for each element. Note that compared with traditional relays, an ARIS performs element-wise amplification/reflection via a diagonal coefficient structure. This can be implemented with much lower hardware and processing overhead, and avoids the spectral efficiency loss in half-duplex relaying. Yet the active amplification of an ARIS also introduces additional noise and is subject to a surface power constraint, necessitating a careful tradeoff in the system design.

\subsection{Signal Model}
Let $x_S \in \mathbb{C}$ be the unit-power transmit signal from $S$, and $x_J \in \mathbb{C}$ be the unit-power jamming signal from $J$. The source $S$ employs a transmit beamforming vector $\bm{w}_S \in \mathbb{C}^{N_S \times 1}$ subject to a maximum transmit power constraint $\|\bm{w}_S\|^2 \le P_S^{\mathrm{max}}$. Also, the jammer $J$ utilizes a transmit beamforming vector $\bm{w}_J \in \mathbb{C}^{N_J \times 1}$ with a power constraint $\|\bm{w}_J\|^2 \le P_J^{\mathrm{max}}$. The received signal $\bm{y}_D \in \mathbb{C}^{N_D \times 1}$ at the destination $D$ is a superposition of signals through the direct and ARIS-reflected paths, and is given by
\begin{align}
\label{eq:yd}
    \bm{y}_D = & \underbrace{(\bm{H}_{RD} \bm{\Theta} \bm{H}_{SR} + \bm{H}_{SD}) \bm{w}_S x_S}_{\text{Desired signal}} \nonumber \\
               & + \underbrace{(\bm{H}_{RD} \bm{\Theta} \bm{H}_{JR} + \bm{H}_{JD}) \bm{w}_J x_J}_{\text{Jamming signal}} \nonumber \\
               & + \underbrace{\bm{H}_{RD} \bm{\Theta} \bm{n}_R}_{\text{Amplified RIS noise}} + \underbrace{\bm{n}_D}_{\text{Receiver noise}},
\end{align}
where $\bm{n}_R \sim \mathcal{CN}(\bm{0}, \sigma_R^2 \bm{I}_N)$ is the noise introduced by the ARIS elements, and $\bm{n}_D \sim \mathcal{CN}(\bm{0}, \sigma_D^2 \bm{I}_{N_D})$ is the additive white Gaussian noise at the receiver. For notation simplicity, we define the effective channels as the cascaded incident and reflection along with the direct transmission as
\begin{align}
    \bm{H}_{SRD} &\triangleq \bm{H}_{RD}\bm{\Theta}\bm{H}_{SR}, &
    \bm{H}_{JRD} &\triangleq \bm{H}_{RD}\bm{\Theta}\bm{H}_{JR}, \\
    \tilde{\bm{H}}_{SD} &\triangleq \bm{H}_{SRD} + \bm{H}_{SD}, &
    \tilde{\bm{H}}_{JD} &\triangleq \bm{H}_{JRD} + \bm{H}_{JD}.
\end{align}
While the destination $D$ employs a receive beamforming vector $\bm{w}_D \in \mathbb{C}^{N_D \times 1}$ with $\|\bm{w}_D\|^2 = 1$, the signal-to-interference-plus-noise ratio (SINR) at the destination is obtained as
\begin{equation} \label{eq:sinr_perfect_csi}
    \Gamma = \frac{|\bm{w}_D^\dagger\tilde{\bm{H}}_{SD}\bm{w}_{S}|^2}
         {|\bm{w}_D^\dagger\tilde{\bm{H}}_{JD}\bm{w}_{J}|^2
         + \sigma_R^2\|\bm{w}_D^\dagger\bm{H}_{RD}\bm{\Theta}\|^2
         + \sigma_D^2}.
\end{equation}
The achieved SINR is a fundamental performance metric directly related to transmission data rate and communication reliability, which can be exploited as a suitable basis for the performance evaluation for both legitimate and adversary sides involved in the system.

\subsection{Uncertainty Model}
For practical communication scenarios, it can be rather challenging to obtain the perfect channel state information due to the random nature of propagation environment. This issue becomes more intricate with considering the anti-jamming communications due to the adversarial relationship between the legitimate side and jammer. In this paper, we consider the case that the jammer has perfect information while the legitimate side only has partial information. This corresponds to the disadvantaged and thus more challenging scenario for the jamming mitigation and also allows effective margins for security provisioning.

Particularly, we assume that the legitimate system has perfect knowledge of its own channels, i.e., $\bm{H}_{SD}$, $\bm{H}_{SR}$, $\bm{H}_{RD}$, but only imperfect estimates of the jamming channels $\bm{H}_{JR}$ and $\bm{H}_{JD}$, i.e., the jammer-related channels. We adopt the bounded error model, where the jamming channels from the legitimate-side perspective are modeled as
\begin{align}
	\bm{H}_{JD} &= \hat{\bm{H}}_{JD} + \triangle\bm{H}_{JD}, \quad \|\triangle\bm{H}_{JD}\|_F^2 \leq \epsilon_{JD}, \label{eq:err-hjd}\\
    \bm{H}_{JR} &= \hat{\bm{H}}_{JR} + \triangle\bm{H}_{JR}, \quad \|\triangle\bm{H}_{JR}\|_F^2 \leq \epsilon_{JR}, \label{eq:err-hjr}
\end{align}
where $\hat{\bm{H}}_{JR}$ and $\hat{\bm{H}}_{JD}$ are the known estimated channel matrices, and $\triangle\bm{H}_{JR}$ and $\triangle\bm{H}_{JD}$ are the corresponding channel estimation errors. These error matrices are measured by the Frobenius norms with bounds $\epsilon_{JR}$ and $\epsilon_{JD}$, respectively. The rationale for this uncertainty configuration includes two folds. First, the legitimate links connect cooperative nodes where the standard pilot-based training can be employed to achieve high-precision channel estimation. In contrast, the jammer is non-cooperative, and its associated channels must be acquired through sensing or blind estimation, and thus more evident errors are unavoidable. Second, we model the errors directly over each jamming link rather than a simplified cascaded aggregation. In this respect, it allows the robust design to strictly capture the interaction between the active reflection and the channel uncertainty, which also ensures that the ARIS effectively mitigates the worst-case jamming impact propagated through the reflection channels. In practice, the uncertainty radii $\epsilon_{JR}$ and $\epsilon_{JD}$ can be set based on the estimation quality, e.g., from the minimum mean square error estimator or from the empirical residual variance of repeated interference sensing snapshots under known reflection configurations.

\section{Game Formulation and Equilibrium} \label{sec:game}

Due to the natural conflicting interest between the legitimate side and adversary, we model their interactions in a Stackelberg game framework, which captures the inherent hierarchical structure in the game playing. We particularly consider the uncertainties from the perspective of legitimate side and formulate the robust Stackelberg game. Then, the equilibrium condition and properties are analyzed to facilitate the strategy design for the game players.

\subsection{Robust Stackelberg Game}

For practical anti-jamming scenarios, the legitimate side usually initiates the transmissions, followed by the jamming attacks. The sequential decision-making process leads to the Stackelberg game model that the legitimate side takes action first as the leader in the game, and the jammer acts sequentially as the follower. In this regard, the leader/follower terminology characterizes the timing and information pattern in the sequential interaction rather than any status hierarchy. For the game players, we introduce the utility functions that characterize their interest as follows. Particularly, for the jammer, as the follower in the game, intending to downgrade or even interrupt the legitimate communications, we employ the negative of achieved SINR as its gain while incorporating the jamming power as the cost to be balanced. Accordingly, we have
\begin{equation} \label{eq:u-j}
    u_J(\bm{w}_J; \bm{w}_S,\bm{w}_D,\bm{\Theta}) = -\Gamma - c_J\|\bm{w}_J\|^2,
\end{equation}
where $c_J > 0$ is the jammer's power consumption cost coefficient, and we explicitly incorporate the legitimate strategy as the function argument to show that the achived utility is affected by both players. Here, we consider the exact SINR, in accordance with the assumption that the jammer has perfect information in the game. Then, the jammer as the follower maximizes its utility function by optimizing the jamming beamformer as
\begin{IEEEeqnarray} {cl} \label{eq:prob-j}
    \IEEEyesnumber \IEEEyessubnumber*
    \max_{\bm{w}_J } & \quad  u_J(\bm{w}_J; \bm{w}_S,\bm{w}_D,\bm{\Theta}) \\
    \rm{s.t.} & \quad \| \bm{w}_J \|^2 \le P_J^{\max}.
\end{IEEEeqnarray}

For the legitimate side as the leader in the game, it intends to achieve secure and reliable transmissions, which is measured by the SINR. However, due to the channel uncertainties from the legitimate perspective, it has no knowledge of the exact SINR. As such, the worst-case estimate of the SINR is adopted to achieve robust security. Also, the power consumption is considered as the cost evaluation. Thus, the utility of the legitimate side is given as
\begin{equation}
    u_L(\bm{w}_S,\bm{w}_D,\bm{\Theta};\bm{w}_J) = \min_{\bm{H}_J\in\mathcal{H}_J} \Gamma-c_S\| \bm{w}_S \|^2,
\end{equation}
where $ c_S $ is the power cost coefficient, $ \bm{H}_J = [\bm{H}_{JR}, \bm{H}_{JD}] $ collects the uncertain channel states and
\begin{equation}
    \mathcal{H}_J = \left\{ \bm{H}_J \left| \begin{aligned} \| \bm{H}_{JR} - \hat{\bm{H}}_{JR} \|_F^2 \leq \epsilon_{JR}, \\ \| \bm{H}_{JD} - \hat{\bm{H}}_{JD} \|_F^2 \leq \epsilon_{JD} \end{aligned} \right. \right\},
\end{equation}
corresponds to the uncertainties defined in~(\ref{eq:err-hjd}) and~(\ref{eq:err-hjr}). Then, the goal of the legitimate side is to maximize its own utility, subject to the transmission and reflection constraints, while considering the uncertainties. Here, we can see that the worst case can be interpreted by two folds. First, for any legitimate strategy, the jammer selects its jamming beamformer and power aiming to maximally degrade the legitimate transmissions. Second, the legitimate side has imperfect knowledge of the jammer-related channels, then the most detrimental channel realization within the uncertainty set needs to be addressed for the legitimate strategy optimization. Accordingly, the legitimate-side problem formulation is given as
\begin{IEEEeqnarray} {cl} \label{eq:prob-l}
    \IEEEyesnumber \IEEEyessubnumber*
    \max_{\bm{w}_S,\bm{w}_D,\bm{\Theta}} & \quad  u_L(\bm{w}_S,\bm{w}_D,\bm{\Theta};\bm{w}_J) \label{eq:l-obj} \\
    \rm{s.t.} & \quad \| \bm{w}_S \|^2 \le P_S^{\max},\quad \|\bm{w}_D\|^2 = 1, \label{eq:pwr-s-con}\\
    &\begin{aligned}\quad \|\bm{\Theta}\bm{H}_{SR}\bm{w}_S\|^2  &+ \max\limits_{\bm{H}_J\in\mathcal{H}_J}\|\bm{\Theta}\bm{H}_{JR}\bm{w}_J\|^2 \\& + \sigma_R^2\|\bm{\Theta}\|_F^2 \leq P_R^{\max}, \end{aligned} \label{eq:pwr-r-con}\\
    % &\quad \|\bm{w}_D\|^2 = 1,\|\bm{w}_S\|^2 = 1,\|\bm{w}_J\|^2 = 1, \\
    &\quad |\bm{\Theta}_{n,n}| \leq \lambda^{\max},\quad\forall n\in \mathcal{N}, \label{eq:refl-con}\\
    &\quad \bm{w}_J^\star = \arg\max_{\bm{w}_J}  u_J(\bm{w}_J; \bm{w}_S,\bm{w}_D,\bm{\Theta}). \label{eq:br-j-con}
\end{IEEEeqnarray}
For the formulated problem, the objective function in~(\ref{eq:l-obj}), as noted before, corresponds to the worst-case legitimate transmission performance, the constraint in~(\ref{eq:pwr-s-con}) specifies the transceiving beamforming power limits, the constraint in~(\ref{eq:pwr-r-con}) denotes the ARIS power budget, which is also affected by the uncertain information, the constraint in{}~(\ref{eq:br-j-con}) indicates that the follower's best response needs to be considered to determine the legitimate strategy.

Therefore, we have introduced the utility functions for the players in the formulated game, where the uncertainty treatment at the legitimate side leads to the robust Stackelberg game model towards robust anti-jamming performance.

\subsection{Equilibrium Analysis}

For a game model, the equilibrium is defined as the strategy profile that no player will unilaterally deviate if other players remain at the equilibrium. For the considered anti-jamming communication scenario, it represents the steady state that both the legitimate side and jammer keep their current actions. Denote the equilibrium strategy profile as $ \left\{ \bm{w}_S^\star,\bm{w}_D^\star,\bm{\Theta}^\star, \bm{w}_J^\star \right\} $, then it holds that
\begin{align}
    u_J(\bm{w}_J^\star; \bm{w}_S^\star,\bm{w}_D^\star,\bm{\Theta}^\star) \ge u_J(\bm{w}_J; \bm{w}_S^\star,\bm{w}_D^\star,\bm{\Theta}^\star),  \label{eq:eq-j}  \\
    \begin{aligned}
    u_L(\bm{w}_S^\star,\bm{w}_D^\star,\bm{\Theta}^\star, \bm{w}_J^\star(\bm{w}_S^\star,\bm{w}_D^\star,\bm{\Theta}^\star)) \qquad\qquad\:\:\\  \ge u_L(\bm{w}_S,\bm{w}_D,\bm{\Theta},\bm{w}_J^\star(\bm{w}_S,\bm{w}_D,\bm{\Theta})),  \label{eq:eq-l} \end{aligned} 
\end{align}
for any other strategy profile $ \left\{ \bm{w}_S,\bm{w}_D,\bm{\Theta}, \bm{w}_J \right\} $ that satisfies the constraints in~(\ref{eq:prob-j}) and~(\ref{eq:prob-l}), for the jammer and legitimate side, respectively. In accordance with the sequential decision-making procedure of the Stackelberg game, the optimal jamming reaction in~(\ref{eq:eq-j}), explicitly represented as~(\ref{eq:br-j-con}), is also incorporated in the leader's optimality condition in~(\ref{eq:eq-l}).

We first analyze the existence of the game equilibrium. Based on the game theories~\cite{E,HanGameBook}, a Stackelberg game admits the equilibrium if the follower's problem has a unique solution for any strategy of the leader. Following this principle, we analyze the jamming strategy determination problem in~(\ref{eq:prob-j}) as follows. To facilitate the analysis, we separate the jamming beamformer as the power and unit-norm direction, denoted by $ P_J $ and $ \hat{\bm{w}}_J $ with $ {\bm{w}}_J = \sqrt{P_J}\hat{\bm{w}}_J $. Hereinafter, we abuse the notation $ \bm{w}_J $ as the unit-norm beamformer while dropping $ \hat{\bm{w}}_J $ for notation simplicity without ambiguity. In this regard, the jammer's optimization becomes
\begin{IEEEeqnarray} {cl} \label{eq:prob-j-alt}
    \IEEEyesnumber \IEEEyessubnumber*
    \max_{P_J, \bm{w}_J } & \quad  u_J(P_J, \bm{w}_J) = -\Gamma-c_J P_J \label{eq:prob-j-alt-obj} \\
    \rm{s.t.} & \quad 0\le P_J \le P_J^{\max}, \quad \| \bm{w}_J \|^2 = 1,
\end{IEEEeqnarray}
where the SINR is re-organized as
\begin{equation} \label{eq:sinr-j}
    \Gamma = \frac{|\bm{w}_D^\dagger\tilde{\bm{H}}_{SD}\bm{w}_{S}|^2}
         {P_J|\bm{w}_D^\dagger\tilde{\bm{H}}_{JD}\bm{w}_{J}|^2
         + \sigma_R^2\|\bm{w}_D^\dagger\bm{H}_{RD}\bm{\Theta}\|^2
         + \sigma_D^2}.
\end{equation}
based on the separate treatment of the jamming power and beamforming direction.

For the problem in (\ref{eq:prob-j-alt}), we investigate the optimization variables individually. For the beamforming direction by $ \bm{w}_J $, it appears in the achieved SINR in~(\ref{eq:sinr-j}), for which the maximization of the utility leads to the minimization of the SINR. Evidently, this problem corresponds to the maximum ratio combining within the denominator of the SINR, leading to the solution as
\begin{equation} \label{eq:opt-j-bf}
    \bm{w}_J^\star = \frac{\tilde{\bm{H}}_{JD}^\dagger\bm{w}_D}
                         {\|\tilde{\bm{H}}_{JD}^\dagger\bm{w}_D\|}.
\end{equation}
Note the jamming beamforming in~(\ref{eq:opt-j-bf}) is not affected by the jamming power but only needs to align with the effective jamming channel. Meanwhile, we can prove that the jammer's utility is a concave function with respect to the jamming power by analyzing the second-order derivative. Then, by nulling the first-order differential of the utility function, we can obtain the optimal jamming power as
\begin{equation} \label{eq:opt-j-pwr}
    P_J^\star = \left[ \frac{|\bm{w}_D^\dagger\tilde{\bm{H}}_{SD}\bm{w}_{S}|}
        {\sqrt{c_J}\|\bm{w}_D^\dagger\tilde{\bm{H}}_{JD}\|}
    -\frac{\sigma_R^2\|\bm{w}_D^\dagger\bm{H}_{RD}\bm{\Theta}\|^2+ \sigma_D^2}
        {\|\bm{w}_D^\dagger\tilde{\bm{H}}_{JD}\|^2} \right]_0^{P_J^{\max}},
\end{equation}
for which we also substitute the optimal jamming beamforming in~(\ref{eq:opt-j-bf}) and $(\cdot)_0^{P_J^{\max}} = \min\{\max\{0, \cdot\}, P_J^{\max}\}$ denotes projection onto the allowed power interval $[0, P_J^{\max}]$.

\section{Robust RIS-Assisted Jamming Mitigation} \label{sec:opt}

Based on the equilibrium analysis above, we have arrived at the jammer's optimal strategy in~(\ref{eq:opt-j-bf}) and~(\ref{eq:opt-j-pwr}), i.e., the analytical results of expression in~(\ref{eq:eq-j}), as part of the equilibrium condition. Then, we solve the leader's problem to obtain the solution to~(\ref{eq:eq-l}) to further constitute the overall equilibrium. To address this issue, we need to jointly consider the legitimate transmission and receiving beamforming as well as the active reflection, in the presence of channel uncertainties to achieve robust anti-jamming communications.

\subsection{Uncertainty Treatment and Problem Decomposition}

Technically, the Stackelberg equilibrium derivation follows the backward induction method, where the follower's best response is first obtained and substituted in the leader's problem to devise the optimality condition at the leader. Following this principle, we substitute the jamming policy in~(\ref{eq:opt-j-bf}) and~(\ref{eq:opt-j-pwr}) to the leader's jamming mitigation problem. To facilitate the discussions of the legitimate strategy, we also extract the legitimate transmit power and the unit-norm beamforming direction as $ P_S $ and $ \bm{w}_S $, respectively, with $ P_S \le P_S^{\max} $ and $ \left\|\bm{w}_S\right\|=1  $, where the notation $ \bm{w}_S $ is reused without loss of ambiguity. Accordingly, the leader's problem becomes
\begin{IEEEeqnarray} {cl} \label{eq:prob-r}
    \IEEEyesnumber \IEEEyessubnumber*
    \max_{\substack{\bm{w}_S,\bm{w}_J,\bm{w}_D,\\P_S,P_J,\bm{\Theta}} }
    \quad& \frac{\sqrt{c_JP_S}|\bm{w}_D^\dagger\tilde{\bm{H}}_{SD}\bm{w}_{S}|}
    {\max\limits_{\bm{H}_J\in\mathcal{H}_J} \|\bm{w}_D^\dagger\tilde{\bm{H}}_{JD}\|}-c_SP_S  \label{eq:r-obj}\\
    \mathrm{s.t.}\quad&0\le P_S\leq P_S^{\max}, \quad 0\le P_J\leq P_J^{\max} , \label{eq:r-pwr-sj}  \\
    &\|\bm{w}_D\|^2 = 1,\quad \|\bm{w}_S\|^2 = 1 ,  \label{eq:r-beam-sd} \\
    &\begin{aligned} P_S\|\bm{\Theta}\bm{H}_{SR}\bm{w}_S\|^2  &+ \max\limits_{\bm{H}_J\in\mathcal{H}_J}P_J\|\bm{\Theta}\bm{H}_{JR}\bm{w}_J\|^2 \\& + \sigma_R^2\|\bm{\Theta}\|_F^2 \leq P_R^{\max}, \end{aligned} \label{eq:r-pwr-r}\\
    & |\bm{\Theta}_{n,n}| \leq \lambda^{\max},\quad\forall n\in \mathcal{N}, \label{eq:r-refl}\\
    & \bm{w}_J^\star = \frac{\tilde{\bm{H}}_{JD}^\dagger\bm{w}_D}{\|\tilde{\bm{H}}_{JD}^\dagger\bm{w}_D\|}, \label{eq:r-beam-j} \\
     & \begin{aligned} P_J \geq  &\frac{\sqrt{P_S}|\bm{w}_D^\dagger\tilde{\bm{H}}_{SD}\bm{w}_{S}|}
        {\min\limits_{\bm{H}_J\in\mathcal{H}_J}\sqrt{c_J}\|\bm{w}_D^\dagger\tilde{\bm{H}}_{JD}\|} \\
    &-\frac{\sigma_R^2\|\bm{w}_D^\dagger\bm{H}_{RD}\bm{\Theta}\|^2+ \sigma_D^2}{\max\limits_{\bm{H}_J\in\mathcal{H}_J}\|\bm{w}_D^\dagger\tilde{\bm{H}}_{JD}\|^2} ,\end{aligned} \label{eq:r-pwr-j}
\end{IEEEeqnarray}
where the follower's best response is employed to rewrite the objective function, and channel uncertainties are explicitly incorporated in~(\ref{eq:r-obj}),~(\ref{eq:r-pwr-r}), and~(\ref{eq:r-pwr-j}). Particularly, we rewrite the follower's optimal power condition in~(\ref{eq:opt-j-pwr}) with newly introduced optimization variable $ P_J $ such that $ P_J \ge P_J^{\star} $. In this regards, the relaxation relieves the difficulties in the original constraints in equality forms, and the allowed higher jamming power presents a more challenging jamming condition. Notably, this relaxation is tight at the leader optimum, since a larger jamming power only makes the leader constraints more restrictive and does not improve the objective, hence an optimal solution always exists with \eqref{eq:r-pwr-j} holding with equality. Moreover, the uncertainties associated with the jamming-related channels are explicitly considered, and thus we arrive at the constraint in~(\ref{eq:r-pwr-j}).

For the problem in~(\ref{eq:prob-r}), we can see that the channel uncertainties appear in the objective function as well as the constraints, which significantly complicates the problem solving. Towards this issue, we first tackle the uncertainties and reformulate the robust conditions in explicit expressions. Particularly, we introduce the auxiliary variables $ \left\{ \psi_{JD}, \phi_{JD}, \psi_{JR} \right\} $  with the conditions
\begin{align}
    \max\limits_{\bm{H}_J\in\mathcal{H}_J}\|\bm{w}_D^\dagger(\bm{H}_{JD}+\bm{H}_{RD}\bm{\Theta}\bm{H}_{JR})\|^2 \,\leq\,  \psi_{JD}, \label{eq:psi-jd} \\
    \min\limits_{\bm{H}_J\in\mathcal{H}_J}\|\bm{w}_D^\dagger(\bm{H}_{JD}+\bm{H}_{RD}\bm{\Theta}\bm{H}_{JR})\|^2 \,\geq\,  \phi_{JD}, \label{eq:phi-jd}  \\
    \max\limits_{\bm{H}_J\in\mathcal{H}_J}\|\bm{\Theta}\bm{H}_{JR}\bm{w}_J\|^2 \,\leq\,  \psi_{JR}, \label{eq:psi-jr}
\end{align}
where the effective channels associated with the jammer are extended to facilitate the uncertainty treatment at the direct and reflection channels. By using the conditions above, the problem in~(\ref{eq:prob-r}) is rewritten as
\begin{IEEEeqnarray} {cl} \label{eq:prob-r-re}
    \IEEEyesnumber \IEEEyessubnumber*
    \max_{\substack{\bm{w}_S,\bm{w}_J,\bm{w}_D,\\P_S,P_J,\bm{\Theta}, \\ \psi_{JD}, \phi_{JD}, \psi_{JR}} }
    \quad& \frac{\sqrt{c_JP_S}|\bm{w}_D^\dagger\tilde{\bm{H}}_{SD}\bm{w}_{S}|}
    {\sqrt{\psi_{JD}}}-c_SP_S \\
    \mathrm{s.t.}\quad&  (\ref{eq:r-pwr-sj}) ,(\ref{eq:r-beam-sd}), (\ref{eq:r-refl}), (\ref{eq:r-beam-j}) \nonumber \\
    &\begin{aligned} &P_S\|\bm{\Theta}\bm{H}_{SR}\bm{w}_S\|^2  + P_J\psi_{JR} \\& \qquad\qquad\qquad+ \sigma_R^2\|\bm{\Theta}\|_F^2 \leq P_R^{\max}, \end{aligned} \label{eq:r-pwr-r-re}\\
     & \begin{aligned} P_J \geq  &\frac{\sqrt{P_S}|\bm{w}_D^\dagger\tilde{\bm{H}}_{SD}\bm{w}_{S}|}
        {\sqrt{c_J}\sqrt{\phi_{JD}}} \\
    &-\frac{\sigma_R^2\|\bm{w}_D^\dagger\bm{H}_{RD}\bm{\Theta}\|^2+ \sigma_D^2}{\psi_{JD}},\end{aligned} \label{eq:r-pwr-j-re} \\
    & (\ref{eq:psi-jd}),(\ref{eq:phi-jd}),(\ref{eq:psi-jr}), \nonumber
\end{IEEEeqnarray}
as a lower-bounded counterpart to the original problem with cleaner expressions.

Meanwhile, the bounded uncertainty conditions in~(\ref{eq:psi-jd}),~(\ref{eq:phi-jd}), and~(\ref{eq:psi-jr}) are tackled as follows. Particularly, for the constraint in~(\ref{eq:psi-jd}), we extend the norm square operation and exploit the Schur complement equivalence with separated channel estimations and errors, and arrive the condition in~(\ref{eq:psi-jd-ext})
\begin{figure*}
\begin{equation} \label{eq:psi-jd-ext}
        \begin{bmatrix}
            \psi_{JD}\bm{I}_{N_J\times N_J}& (\bm{w}_D^\dagger\hat{ \bm{H}}_{JD}+\bm{w}_D^\dagger\bm{H}_{RD}\bm{\Theta}\hat{\bm{H}}_{JR})^\dagger\\
            \bm{w}_D^\dagger\hat{\bm{H}}_{JD}+\bm{w}_D^\dagger\bm{H}_{RD}\bm{\Theta}\hat{\bm{H} }_{JR} & 1 \\
        \end{bmatrix}
       \succcurlyeq \bm{\Xi}_{JD} + \bm{\Xi}^{\dag}_{JD} + \bm{\Xi}_{JR} + \bm{\Xi}^{\dag}_{JR}
\end{equation}
% \hrulefill
\end{figure*}
with
\begin{equation}
	\bm{\Xi}_{JD} = -\begin{bmatrix}
            \bm{I}_{N_J\times N_J}\\
            \bm{0}_{1\times N_J}
        \end{bmatrix}\triangle\bm{H}_{JD}^\dagger
        \begin{bmatrix}
            \bm{0}_{N_D\times N_J}& \bm{w}_D
        \end{bmatrix},
\end{equation}
and
\begin{equation}
	\bm{\Xi}_{JR} = -\begin{bmatrix}
            \bm{I}_{N_J\times N_J}\\
            \bm{0}_{1\times N_J}
        \end{bmatrix}\triangle\bm{H}_{JR}^\dagger
        \begin{bmatrix}
            \bm{0}_{N\times N_J}& (\bm{w}_D^\dagger\bm{H}_{RD}\bm{\Theta})^\dagger
        \end{bmatrix}.
\end{equation}
\begin{figure*}
\begin{equation}\label{eq:psi-jd-mat}
	\begin{aligned}
    \begin{bmatrix}
        (\psi_{JD}-\rho_1-\rho_2)\bm{I}_{N_J\times N_J}&(\bm{w}_D^\dagger\hat{ \bm{H}}_{JD}+\bm{w}_D^\dagger\bm{H}_{RD}\bm{\Theta}\hat{\bm{H}}_{JR})^\dagger
            &\bm{0}_{N_J\times N_D}&\bm{0}_{N_J\times N}\\
        \bm{w}_D^\dagger\hat{ \bm{H}}_{JD}+\bm{w}_D^\dagger\bm{H}_{RD}\bm{\Theta}\hat{\bm{H}}_{JR}&1
            &\sqrt{\epsilon_{JD}}\bm{w}_D^\dagger & \sqrt{\epsilon_{JR}}\bm{w}_D^\dagger\bm{H}_{RD}\bm{\Theta}\\
        \bm{0}_{N_D\times N_J}&\sqrt{\epsilon_{JD}}\bm{w}_D&\rho_1\bm{I}_{N_D\times N_D}&\bm{0}_{N_D\times N}\\
        \bm{0}_{N\times N_J}&\sqrt{\epsilon_{JR}}(\bm{w}_D^\dagger\bm{H}_{RD}\bm{\Theta})^\dagger&\bm{0}_{N\times N_D}&\rho_2\bm{I}_{N\times N}
    \end{bmatrix}\, \qquad\qquad \\ \succcurlyeq\, \bm{0}_{(N_J+N+N_D+1)\times (N_J+N+N_D+1)}
\end{aligned}
\end{equation}
\hrulefill
\end{figure*}
For the condition in~(\ref{eq:psi-jd-ext}), we can further apply the general sign-definiteness principle~\cite{math1} by considering the bounded uncertainties in $ \mathcal{H} $ and arrive at the condition in~(\ref{eq:psi-jd-mat}), where $ \rho_1 $ and $ \rho_2 $ are the introduced auxiliary variables due to the uncertainty bounds in~(\ref{eq:err-hjd}) and~(\ref{eq:err-hjr}), respectively.
To facilitate the reformulation of condition in~(\ref{eq:phi-jd}), we introduce
\begin{equation}
\bm{b}^\dagger \,\triangleq\,
    \begin{bmatrix}
        \bm{w}_D^\dagger&\bm{w}_D^\dagger\bm{H}_{RD}\bm{\Theta}
    \end{bmatrix},
\end{equation}
and
\begin{equation}
\bm{H}_J^\dagger \,\triangleq\,
    \begin{bmatrix}
        \bm{H}_{JD}\\
        \bm{H}_{JR}
    \end{bmatrix}
    =\underbrace{\begin{bmatrix}
        \hat{\bm{H}}_{JD}\\
        \hat{\bm{H}}_{JR}
    \end{bmatrix}}_{=\hat{\bm{H}}_J^\dagger}
    +\underbrace{\begin{bmatrix}
        \triangle\bm{H}_{JD}\\
        \triangle\bm{H}_{JR}
    \end{bmatrix}}_{=\triangle\bm{H}_J^\dagger},
\end{equation}
and the condition in~(\ref{eq:phi-jd}) is simplified as $ \bm{b}^\dagger\bm{H}_J^\dagger\bm{H}_J\bm{b} \geq \phi_{JD} $. By invoking the trace equality, we can further derive the equivalences as
\begin{equation} \label{eq:def-B}
\begin{aligned}
	&\mathsf{Tr}(\bm{H}_J^\dagger\bm{H}_J\bm{b}\bm{b}^\dagger)-\phi_{JD}\geq0 \\
	\Rightarrow & \mathrm{vec}^\dagger(\bm{H}_J)\bm{B}\mathrm{vec}(\bm{H}_J)-\phi_{JD}\geq0,
\end{aligned}
\end{equation}
with $\bm{B}=(\bm{b}\bm{b}^\dagger)^T\otimes \bm{I}_{N_J\times N_J}$. Then, by separating the channel estimations and errors, we the inequality above can be extended as
\begin{equation} \label{eq:phi-jd-ext}
\begin{aligned}
    \mathrm{vec}^\dagger(\hat{\bm{H}}_J)\bm{B}\mathrm{vec}(\hat{\bm{H}}_J)
        +\mathrm{vec}^\dagger(\hat{\bm{H}}_J)\bm{B}\mathrm{vec}(\triangle\bm{H}_J) \\
        +\mathrm{vec}^\dagger(\triangle\bm{H}_J)\bm{B}\mathrm{vec}(\hat{\bm{H}}_J)
        +\mathrm{vec}^\dagger(\triangle\bm{H}_J)\bm{B}\mathrm{vec}(\triangle\bm{H}_J) \\
        -\phi_{JD}\geq0.
\end{aligned}
\end{equation}
Revisiting the error bounds in~(\ref{eq:err-hjd}) and~(\ref{eq:err-hjr}), we can vectorize the errors in the form of
\begin{equation} \label{eq:upsilon-jd}
	\begin{aligned}
    \mathrm{vec}^\dagger(\triangle\bm{H}_{J})
    \underbrace{\begin{bmatrix}
        \bm{I}_{N_JN_D\times N_JN_D}&\bm{0}_{N_JN_D\times N_JN}\\
        \bm{0}_{N_JN\times N_JN_D}&\bm{0}_{N_JN\times N_JN}
    \end{bmatrix}}_{\triangleq\,\bm{\Upsilon}_{JD}}
    \mathrm{vec}(\triangle\bm{H}_{J}) \\
  	-\epsilon _{JD}\leq0,
    \end{aligned}
\end{equation}
and
\begin{equation} \label{eq:upsilon-jr}
	\begin{aligned}
    \mathrm{vec}^\dagger(\triangle\bm{H}_{J})
    \underbrace{\begin{bmatrix}
        \bm{0}_{N_JN_D\times N_JN_D}&\bm{0}_{N_JN_D\times N_JN}\\
        \bm{0}_{N_JN\times N_JN_D}&\bm{I}_{N_JN\times N_JN}
    \end{bmatrix}}_{\triangleq\,\bm{\Upsilon}_{JR}}
    \mathrm{vec}(\triangle\bm{H}_{J})\\
    -\epsilon _{JR}\leq0.
 \end{aligned}
\end{equation}
By invoking the S-procedure~\cite{math2} to incorporate the error bounds in~(\ref{eq:err-hjd}) and~(\ref{eq:err-hjr}), the inequality in~(\ref{eq:phi-jd-ext}) can be derived in the form of~(\ref{eq:phi-jd-mat}), where $ \eta_1 $ and $ \eta_2 $ are introduced variables associated with~(\ref{eq:upsilon-jd}) and~(\ref{eq:upsilon-jr}), respectively.
\begin{figure*}
	\begin{equation} \label{eq:phi-jd-mat}
    \begin{bmatrix}
        \bm{B}+\eta_1\bm{\Upsilon}_{JD}+\eta_2\bm{\Upsilon}_{JR}&\bm{B}\mathrm{vec}(\hat{\bm{H}}_{J})\\
        \mathrm{vec}^\dagger(\hat{\bm{H}}_{J})\bm{B}
        & \mathrm{vec}^\dagger(\hat{\bm{H}}_J)\bm{B}\mathrm{vec}(\hat{\bm{H}}_J)-\phi_{JD}-\eta_1\epsilon_{JD}-\eta_2\epsilon_{JR}
    \end{bmatrix}\,\succcurlyeq\, \bm{0}_{(N_J(N+N_D)+1)\times (N_J(N+N_D)+1)}
 \end{equation}
\hrulefill
\end{figure*}
Finally, for the constraint in~(\ref{eq:psi-jr}), we can follow the same procedure to tackle the constraint in~(\ref{eq:psi-jd}) by employing the Schur complement and general sign-definiteness principle~\cite{math1}, and thus arrive the error-bounded condition as
\begin{equation} \label{eq:psi-jr-mat}
\begin{aligned}
    \begin{bmatrix}
        \psi_{JR}\bm{I}_{N_J\times N_J}&\bm{\Theta}\hat{\bm{H}}_{JR}\bm{w}_J
            &\sqrt{\epsilon_{JR}}\bm{\Theta} \\
        \bm{w}_J^\dagger\hat{\bm{H}}_{JR}^\dagger\bm{\Theta}^\dagger &1-\rho_3 \bm{w}_J^\dagger\bm{w}_J
             & \bm{0}_{1\times N}\\
        \sqrt{\epsilon_{JR}}\bm{\Theta}^\dagger&\bm{0}_{N\times 1}&\rho_3\bm{I}_{N\times N}
    \end{bmatrix} \qquad\qquad \\ \succcurlyeq\, \bm{0}_{(2N+1)\times (2N+1)}	,
\end{aligned}
\end{equation}
where $ \rho_3 $ is the auxiliary variable associated with the error constraint in~(\ref{eq:err-hjr}). The detailed derivations follow the same procedure as elaborated before and thus are omitted for brevity.

As the channel uncertainties have been tackled with the bound information, then leader's problem in~(\ref{eq:prob-r}) is reformulated as
\begin{IEEEeqnarray} {cl} \label{eq:prob-r-all}
    \IEEEyesnumber \IEEEyessubnumber*
    \max_{\substack{\bm{w}_S,\bm{w}_J,\bm{w}_D,\\P_S,P_J,\bm{\Theta}, \\ \psi_{JD}, \phi_{JD}, \psi_{JR}, \\ \rho_1,\rho_2,\rho_3, \eta_1, \eta_2} }
    \quad& \frac{\sqrt{c_JP_S}|\bm{w}_D^\dagger\tilde{\bm{H}}_{SD}\bm{w}_{S}|}
    {\sqrt{\psi_{JD}}}-c_SP_S  \\
    \mathrm{s.t.}\quad&  (\ref{eq:r-pwr-sj}) ,(\ref{eq:r-beam-sd}), (\ref{eq:r-refl}), (\ref{eq:r-beam-j}), \nonumber \\
    & (\ref{eq:r-pwr-r-re}), (\ref{eq:r-pwr-j-re}), \nonumber \\
    & (\ref{eq:psi-jd-mat}),(\ref{eq:phi-jd-mat}),(\ref{eq:psi-jr-mat}), \nonumber \\
    & \rho_1\ge0,\rho_2\ge0,\rho_3\ge0, \label{eq:rhos}\\
    & \eta_1\ge0, \eta_2\ge0. \label{eq:etas}
\end{IEEEeqnarray}
As compared with its original counterpart, the problem in~(\ref{eq:prob-r-all}) recast the robustness conditions with explicit expressions by employing the matrix inequalities with the additional auxiliary variables.  To solve the problem effectively, we decompose the problem within a BSUM framework\footnote{The BSUM framework is generally used for minimization problem, while we consider a maximization problem, it is actually the block successive lower-bound maximization (BSLM). Yet for the consistency with the mainstream terminology, we still refer to it as BSUM.} to tackle the transmit power, beamforming, and reflection, respectively, and iteratively update blocked optimization variables to approximate the suboptimal solution to the legitimate-side strategy.

\subsection{Power Allocation}

We first consider the subproblem of legitimate transmit power and jamming power allocation while fixing the other optimization variables. In this regard, the subproblem is obtained as
\begin{IEEEeqnarray} {cl} \label{eq:subprob-pwr}
    \IEEEyesnumber \IEEEyessubnumber*
    \max_{P_S,P_J }
    \quad& \frac{\sqrt{c_JP_S}|\bm{w}_D^\dagger\tilde{\bm{H}}_{SD}\bm{w}_{S}|}
    {\sqrt{\psi_{JD}}}-c_SP_S  \\
    \mathrm{s.t.}\quad&  (\ref{eq:r-pwr-sj}),  (\ref{eq:r-pwr-r-re}), (\ref{eq:r-pwr-j-re}), \nonumber
\end{IEEEeqnarray}
where the non-convexity is due to the square root operation of the legitimate transmit power. As such, we introduce a counterpart variable $ \gamma \ge 0 $ such that $ \gamma^2 = P_S $, then the problem is reformulated as
\begin{IEEEeqnarray} {cl} \label{eq:subprob-pwr-equiv}
    \IEEEyesnumber \IEEEyessubnumber*
    \max_{\gamma,P_J }
    \quad& \frac{\sqrt{c_J}\gamma|\bm{w}_D^\dagger\tilde{\bm{H}}_{SD}\bm{w}_{S}|}
    {\sqrt{\psi_{JD}}}-c_S\gamma^2  \\
    \mathrm{s.t.}\quad& 0\le P_J\leq P_J^{\max} ,   \\ 
    & \gamma \ge 0, \quad \gamma^2 \le P_S^{\max}, \\
    &\begin{aligned} &\gamma^2\|\bm{\Theta}\bm{H}_{SR}\bm{w}_S\|^2  + P_J\psi_{JR} \\& \qquad\qquad\qquad+ \sigma_R^2\|\bm{\Theta}\|_F^2 \leq P_R^{\max}, \end{aligned} \\
     & \begin{aligned} P_J \geq  &\frac{\gamma|\bm{w}_D^\dagger\tilde{\bm{H}}_{SD}\bm{w}_{S}|}{\sqrt{c_J}\sqrt{\phi_{JD}}} \\
    &-\frac{\sigma_R^2\|\bm{w}_D^\dagger\bm{H}_{RD}\bm{\Theta}\|^2+ \sigma_D^2}{\psi_{JD}}.\end{aligned}
\end{IEEEeqnarray}
As we can conveniently verify, the problem in~(\ref{eq:subprob-pwr-equiv}) is convex and thus can be readily tackled with off-the-shelf solvers. Denote the obtained solution as $ \left\{ \gamma^\star, P_J^{\star}  \right\} $, then the legitimate transmit power is determined as $ P_S^\star = \left(\gamma^\star\right)^2 $. The obtained power allocation of $ \left\{P_S^\star, P_J^\star\right\} $ are then used for the subproblem solving of other variables in the legitimate-side problem.

\subsection{Beamforming Determination}

For the legitimate side, it needs to determine the transmit and receive beamforming due to the multi-antenna capability. Meanwhile, the jamming beamforming is also considered since the legitimate side as the leader in the game needs to anticipate the jamming reaction at the follower to support the legitimate strategy derivation. We then discuss each of the beamforming vector individually.

The jamming beamforming is determined at the adversary, as explicitly specified in~(\ref{eq:r-beam-j}) based on the best-response analysis. Meanwhile, for the transmit beamforming at the legitimate side, it reduces to the problem as $ \max_{\|\bm{w}_S\|^2 = 1} \quad
     \left| \bm{w}_D^\dagger\tilde{\bm{H}}_{SD}\bm{w}_{S} \right| $.
We can readily see that it is a maximum ratio transmission problem, and thus the optimal transmit beamforming needs to align with the effective channel condition as
\begin{equation} \label{eq:opt-ws}
    \bm{w}_S^\star = \frac{\tilde{\bm{H}}_{SD}^\dagger\bm{w}_D}
                         {\|\tilde{\bm{H}}_{SD}^\dagger\bm{w}_D\|}.
\end{equation}
Note here the legitimate side has the perfect information regarding the legitimate-side channels, which enables the beamforming calculation in~(\ref{eq:opt-ws}). 

When considering the receive beamforming at the legitimate side, the problem becomes
\begin{IEEEeqnarray} {cl} \label{eq:subprob-wd}
    \IEEEyesnumber \IEEEyessubnumber*
    \max_{\substack{\left\| \bm{w}_D \right\|=1,\\ \psi_{JD}, \phi_{JD}, \\ \rho_1,\rho_2, \eta_1, \eta_2} }
    \quad& \frac{|\bm{w}_D^\dagger\tilde{\bm{H}}_{SD}\bm{w}_{S}|}
    {\sqrt{\psi_{JD}}}  \\
    \mathrm{s.t.}\quad& (\ref{eq:r-pwr-j-re}), (\ref{eq:psi-jd-mat}),(\ref{eq:phi-jd-mat}), \nonumber \\
    & (\ref{eq:rhos}),(\ref{eq:etas}).
\end{IEEEeqnarray}
As can be easily verified, the non-convexity lies in the objective function, the unit-norm nature of beamformer, and the constraints in~(\ref{eq:r-pwr-j-re}) and~(\ref{eq:phi-jd-mat}), while the constraints in~(\ref{eq:psi-jd-mat}) is a linear matrix inequality with respect to $ \bm{w}_D $, and the rest constraints are linear. Then, the non-convexity treatments are conducted as follows.

For the unit-norm constraint of the beamforming vector, we introduce $ \bm{W}_D = \bm{w}_D\bm{w}_D^\dagger $, and the unit norm leads to the condition that $ \mathsf{Tr}(\bm{W}_D) = 1 $. By rewriting the equality constraint as $ \bm{W}_D \ge \bm{w}_D\bm{w}_D^\dagger $ and $ \bm{W}_D \le \bm{w}_D\bm{w}_D^\dagger $, we derive the equivalent condition for the unit-norm constraint as
\begin{subnumcases}{\label{eq:Wd-con}}
    \mathsf{Tr}(\bm{W}_D)=1,\\
    \begin{bmatrix}
        \bm{W}_D &\bm{w}_D\\
        \bm{w}_D^\dagger &1
    \end{bmatrix}\succcurlyeq \bm{0}_{(N_D+1)\times (N_D+1)},\\
    \mathsf{Tr}(\bm{W}_D-\bm{w}_D\bm{w}_D^\dagger)\leqslant 0,
\end{subnumcases}
in the form of linear matrix inequalities as the convex counterpart. For the constraint in~(\ref{eq:r-pwr-j-re}), we introduce a set of auxiliary variables $ \left\{ \mu_D, \nu_D, \xi_D \right\} $ such that
\begin{subnumcases}{}
   \mu_D^2 \geq |\bm{w}_D^\dagger\tilde{\bm{H}}_{SD}\bm{w}_{S}|, \label{eq:mu-d}  \\
   \nu_D \leq \sqrt{\phi_{JD}}, \label{eq:nu-d} \\
   \xi_D^2 \leq \sigma_R^2\|\bm{w}_D^\dagger\bm{H}_{RD}\bm{\Theta}\|^2+ \sigma_D^2 ,  \label{eq:xi-d}
\end{subnumcases}
which leads to the lower-bound reformulation of~(\ref{eq:r-pwr-j-re}) as
\begin{equation} \label{eq:pj-d}
	    P_J \geq \frac{\sqrt {P_S}}{\sqrt{c_J}} \frac{\mu_D^2}{\nu_D}- \frac{\xi_D^2}{\psi_{JD}}.
\end{equation}
Here, the auxiliary-related constraints in~(\ref{eq:mu-d}), (\ref{eq:xi-d}), and the induced constraint in~(\ref{eq:pj-d}) are non-convex. In this respect, we can adopt the SCA technique to use the first-order approximated counterpart for the non-convex parts and thus obtain
\begin{equation} \label{eq:mu-d-apx}
    2\mu_{D}^{\circ}\mu_D-(\mu_{D}^{\circ})^2 \geq |\bm{w}_D^\dagger\tilde{\bm{H}}_{SD}\bm{w}_{S}|,
\end{equation} 
and
\begin{equation}\label{eq:nu-d-apx}
\begin{aligned}
	\xi_D^2 \leq \sigma_R^2\mathsf{Tr}\left(\bm{H}_{RD}\bm{\Theta}\bm{\Theta}^\dagger\bm{H}_{RD}^\dagger\left(\bm{w}_D(\bm{w}_{D}^{\circ})^\dagger \right.\right. \quad\\
    \left.\left. \bm{w}_{D}^{\circ}\bm{w}_D^\dagger-\bm{w}_{D}^{\circ}(\bm{w}_{D}^{\circ})^\dagger\right)\right)+ \sigma_D^2,
\end{aligned}
\end{equation}
along with
\begin{equation} \label{eq:pj-d-apx}
	P_J \geq \frac{\sqrt {P_S}}{\sqrt{c_J}} \frac{\mu_D^2}{\nu_D}-\frac{2\xi_{D}^{\circ}\xi_D\psi_{JD}^{\circ}-(\xi_{D}^{\circ})^2\psi_{JD}}{(\psi_{JD}^{\circ})^2},
\end{equation}
as the convexified counterparts at the approximation points $ \left\{ \bm{w}_{D}^{\circ},\mu_{D}^{\circ},  \xi_{D}^{\circ}, \psi_{JD}^{\circ} \right\} $.
For the non-convexity in~(\ref{eq:psi-jd-mat}), it originates from the definition in $\bm{B}$ with the square operation over the receive beamformer $ \bm{w}_D $. To tackle this issue, we trace back to the initial introduction of $\bm{B}$ in~(\ref{eq:def-B}), for which a lower-bounded approximation through SCA is derived as~(\ref{eq:bb-apx}) at the point $ \bm{w}_D^{\circ} $.
\begin{figure*}
\begin{equation} \label{eq:bb-apx}
\bm{b}\bm{b}^\dagger \succcurlyeq 
\underbrace{\begin{bmatrix}
    \bm{w}_D(\bm{w}_{D}^{\circ})^\dagger+\bm{w}_{D}^{\circ}\bm{w}_D^\dagger-\bm{w}_{D}^{\circ}(\bm{w}_{D}^{\circ})^\dagger
    &\left(\bm{w}_D(\bm{w}_{D}^{\circ})^\dagger+\bm{w}_{D}^{\circ}\bm{w}_D^\dagger-\bm{w}_{D}^{\circ}(\bm{w}_{D}^{\circ})^\dagger\right)\bm{H}_{RD}\bm{\Theta}\\
    \bm{\Theta}^\dagger\bm{H}_{RD}^\dagger\left(\bm{w}_D(\bm{w}_{D}^{\circ})^\dagger+\bm{w}_{D}^{\circ}\bm{w}_D^\dagger-\bm{w}_{D}^{\circ}(\bm{w}_{D}^{\circ})^\dagger\right)
    &\bm{\Theta}^\dagger\bm{H}_{RD}^\dagger\left(\bm{w}_D(\bm{w}_{D}^{\circ})^\dagger+\bm{w}_{D}^{\circ}\bm{w}_D^\dagger-\bm{w}_{D}^{\circ}(\bm{w}_{D}^{\circ})^\dagger\right)\bm{H}_{RD}\bm{\Theta}
\end{bmatrix}}_{\triangleq\,\bm{A}_{D} \left(\bm{w}_D; \bm{w}_D^{\circ}\right) }
\end{equation}
\begin{equation} \label{eq:B-apx}
\begin{aligned}
	    \begin{bmatrix}
        \bm{B}_{D}\left(\bm{w}_D; \bm{w}_D^{\circ}\right)+\eta_1\bm{\gamma}_{JD}+\eta_2\bm{\gamma}_{JR}&\bm{B}_{D}\left(\bm{w}_D; \bm{w}_D^{\circ}\right)\mathrm{vec}(\hat{\bm{H}}_{J})\\
        \mathrm{vec}^\dagger(\hat{\bm{H}}_{J})\bm{B}_{D}\left(\bm{w}_D; \bm{w}_D^{\circ}\right)
        & \mathrm{vec}^\dagger(\hat{\bm{H}}_J)\bm{B}_{D}\left(\bm{w}_D; \bm{w}_D^{\circ}\right)\mathrm{vec}(\hat{\bm{H}}_J)-\phi_{JD}-\eta_1\epsilon_{JD}-\eta_2\epsilon_{JR}
    \end{bmatrix} \qquad\qquad \\  \succcurlyeq\, \bm{0}_{(N_J(N+N_D)+1)\times (N_J(N+N_D)+1)}
\end{aligned}
\end{equation}
\hrulefill
\end{figure*}
Then, we define that $\bm{B}_{D}\left(\bm{w}_D; \bm{w}_D^{\circ}\right)=\bm{A}_{D} \left(\bm{w}_D; \bm{w}_D^{\circ}\right)\otimes\bm{I}_{N_J \times N_J}$ to approximate $ \bm{B} $, and the constraint in~(\ref{eq:def-B}) is tightened by $ \mathrm{vec}^\dagger(\bm{H}_J) \bm{B}_{D}\left(\bm{w}_D; \bm{w}_D^{\circ}\right) \mathrm{vec}(\bm{H}_J)-\phi_{JD}\geq0 $. By adopting the uncertainty treatment as extended before, we derive the condition in~(\ref{eq:B-apx}) as a replacement of the non-convex constraint in~(\ref{eq:phi-jd-mat}), in the form of a linear matrix inequality against $ \bm{w}_D $.
Finally, for the non-convex objective function, we similarly introduce $ \left\{ \phi_D, \psi_D  \right\} $ such that
\begin{subnumcases}{}
 \phi_D^2 \leq  |\bm{w}_D^\dagger\tilde{\bm{H}}_{SD}\bm{w}_{S}|, \label{eq:phi-d} \\
 \psi_D \geq  \sqrt{\psi_{JD}}, \label{eq:psi-d}
\end{subnumcases}
which leads the lower bound of the objective function that
\begin{equation} \label{eq:obj-d}
	\frac{|\bm{w}_D^\dagger\tilde{\bm{H}}_{SD}\bm{w}_{S}|} {\sqrt{\psi_{JD}}} \ge \frac{\phi_D^2}{\psi_D}.
\end{equation}
Furthermore, for the reinterpretation of the objective function and the inherent non-convexity, we adopt the SCA technique and arrive the convexified version that
\begin{subnumcases}{\label{eq:phi-psi-d-apx}}
  \begin{aligned}
  \phi_D^4 \leq  \bm{w}_S^\dagger\tilde{\bm{H}}_{SD}^\dagger(\bm{w}_D(\bm{w}_{D}^{\circ})^\dagger+\bm{w}_{D}^{\circ}\bm{w}_D^\dagger \qquad \\
  -\bm{w}_{D}^{\circ}(\bm{w}_{D}^{\circ})^\dagger)\tilde{\bm{H}}_{SD}\bm{w}_{S} ,	
  \end{aligned}
  \\
  2\psi_D\psi_D^{\circ} - (\psi_D^{\circ})^2 \geq  \psi_{JD},
\end{subnumcases}
along with
\begin{equation}
	\frac{\phi_D^2}{\psi_D} \geq \frac{2\phi_D\phi_{D}^{\circ}\psi_{D}^{\circ}-(\phi_{D}^{\circ})^2\psi_D}{(\psi_{D}^{\circ})^2},
\end{equation}
as approximated at $ \left\{ \bm{w}_D^{\circ}, \psi_D^{\circ}, \phi_D^{\circ}  \right\} $.

Therefore, by tackling the non-convexity in the receive beamforming optimization in~(\ref{eq:subprob-wd}), we arrive a lower-bounded surrogate problem in the form of
\begin{IEEEeqnarray} {cl} \label{eq:subprob-wd-apx}
    \IEEEyesnumber \IEEEyessubnumber*
    \max_{\substack{\bm{w}_D ,\psi_{JD}, \phi_{JD}, \\ \rho_1,\rho_2, \eta_1, \eta_2, \\ \mu_D,\nu_D,\xi_D, \phi_D, \psi_D} }
    \quad& \frac{2\phi_D\phi_{D}^{\circ}\psi_{D}^{\circ}-(\phi_{D}^{\circ})^2\psi_D}{(\psi_{D}^{\circ})^2}  \\
    \mathrm{s.t.}\quad&  (\ref{eq:psi-jd-mat}),(\ref{eq:rhos}),(\ref{eq:etas}), \nonumber \\
    & (\ref{eq:Wd-con}),(\ref{eq:nu-d}) ,(\ref{eq:mu-d-apx}),   (\ref{eq:nu-d-apx}),(\ref{eq:pj-d-apx}),(\ref{eq:B-apx}), (\ref{eq:phi-psi-d-apx}), \nonumber
\end{IEEEeqnarray}
which is now a convex optimization approximated at $ \left\{ \bm{w}_{D}^{\circ},\mu_{D}^{\circ},  \xi_{D}^{\circ}, \psi_{JD}^{\circ}, \psi_D^{\circ}, \phi_D^{\circ} \right\} $. Then, we solve a series of problems in the form of~(\ref{eq:subprob-wd-apx}), and the use the obtained optimum to update the approximation points, where the convergence corresponds to the solution to the receive beamforming optimization in~(\ref{eq:subprob-wd}).

\subsection{Reflection Optimization}

The last subproblem through the BSUM decomposition addresses the reflection optimization at the ARIS, which determines the amplification ratio and reflection phase, while fixing the transmit powers and beamforming vectors. Accordingly, the reflection optimization is in the form of
\begin{IEEEeqnarray} {cl} \label{eq:subprob-Theta}
    \IEEEyesnumber \IEEEyessubnumber*
    \max_{\substack{ \bm{\Theta} ,\\ \psi_{JD}, \psi_{JR}, \phi_{JD}, \\ \rho_1,\rho_2,\rho_3, \eta_1, \eta_2 } }
    \quad& \frac{|\bm{w}_D^\dagger\tilde{\bm{H}}_{SD}\bm{w}_{S}|}
        {\sqrt{\psi_{JD}}}  \\
        \mathrm{s.t.}\quad&  (\ref{eq:r-refl}),  (\ref{eq:r-pwr-r-re}), (\ref{eq:r-pwr-j-re}), \nonumber \\
    & (\ref{eq:psi-jd-mat}),(\ref{eq:phi-jd-mat}),(\ref{eq:psi-jr-mat}), \nonumber \\
    & (\ref{eq:rhos}),(\ref{eq:etas}). \nonumber
\end{IEEEeqnarray}
Since the reflection matrix has a high dimension while only the diagonal elements are effective, we extract the diagonal vector to facilitate the analysis. To this end, conduct the transformation that
\begin{equation}
    \begin{aligned}
    \bm{w}_D^\dagger\tilde{\bm{H}}_{SD}\bm{w}_{S}
    = &\bm{w}_D^\dagger(\bm{H}_{RD}\bm{\Theta}\bm{H}_{SR}+\bm{H}_{SD})\bm{w}_{S} \\
    = &\underbrace{\bm{w}_D^\dagger\bm{H}_{RD}\mathsf{diag}(\bm{H}_{SR}\bm{w}_{S})}_{\triangleq \bm{\alpha}^\dagger}\bm{\theta}
    +\underbrace{\bm{w}_D^\dagger\bm{H}_{SD}\bm{w}_{S}}_{\triangleq\beta} \\
    = & \bm{\alpha}^\dagger\bm{\theta} + \beta  ,
\end{aligned} 
\end{equation}
and
\begin{equation}
    \bm{w}_D^\dagger\bm{H}_{RD}\bm{\Theta}
    = \bm{\theta}^\dagger \underbrace{\mathsf{diag}(\bm{w}_D^\dagger\bm{H}_{RD})}_{\triangleq  \bm{E}}
    = \bm{\theta}^\dagger\bm{E},
\end{equation}
along with the equality that
\begin{subnumcases}{\label{eq:Theta-theta}}
    \bm{\Theta}\bm{H}_{SR}\bm{w}_S = \mathsf{diag}(\bm{H}_{SR}\bm{w}_S)\bm{\theta}, \\
    \bm{\Theta}\bm{H}_{JR} = \mathsf{diag}(\bm{\theta})\bm{H}_{JR}, \\
    \|\bm{\Theta}\|_F^2 = \|\bm{\theta}\|^2.
\end{subnumcases}
Then, the problem in~(\ref{eq:subprob-Theta}) can be rewritten with the optimization of the reflection vector $ \bm{\theta} $ rather than the reflection matrix $ \bm{\Theta} $, specified as
\begin{IEEEeqnarray} {cl} \label{eq:subprob-theta}
    \IEEEyesnumber \IEEEyessubnumber*
    \max_{\substack{ \bm{\theta} ,\\ \psi_{JD}, \psi_{JR}, \phi_{JD}, \\ \rho_1,\rho_2,\rho_3, \eta_1, \eta_2 } }
    \quad& \frac{\left| \bm{\alpha}^\dagger\bm{\theta} + \beta \right|}{\sqrt{\psi_{JD}}}  \label{eq:theta-obj} \\
        \mathrm{s.t.}\quad&  |{\theta}_{n}| \leq \lambda^{\max},\quad\forall n\in \mathcal{N}, \label{eq:theta-n}\\
    &\begin{aligned} &P_S\| \mathsf{diag}(\bm{H}_{SR}\bm{w}_S)\bm{\theta} \|^2  + P_J\psi_{JR} \\& \qquad\qquad+ \sigma_R^2\|\bm{\theta}\|^2 \leq P_R^{\max}, \end{aligned} \label{eq:pwr-r-theta}\\
     & \begin{aligned} P_J \geq  &\frac{\sqrt{P_S}\left| \bm{\alpha}^\dagger\bm{\theta} + \beta \right|}
        {\sqrt{c_J}\sqrt{\phi_{JD}}} \\
    &-\frac{\sigma_R^2\| \bm{\theta}^\dagger\bm{E} \|^2+ \sigma_D^2}{\psi_{JD}},\end{aligned} \label{eq:pwr-j-theta} \\
    & (\ref{eq:psi-jd-mat}),(\ref{eq:phi-jd-mat}),(\ref{eq:psi-jr-mat}), \nonumber \\
    & (\ref{eq:rhos}),(\ref{eq:etas}), \nonumber
\end{IEEEeqnarray}
where for the constraints in~(\ref{eq:psi-jd-mat}),~(\ref{eq:phi-jd-mat}),~(\ref{eq:psi-jr-mat}), $ \bm{\Theta} $ is replaced by $\mathsf{diag}(\bm{\theta}) $, and we omit the specified expressions and reuse the equation labels for space limitation. For the reformulated reflection optimization in~(\ref{eq:subprob-theta}), we can see that the reflection amplitude constraint in~(\ref{eq:theta-n}) and power constraint at the ARIS in~(\ref{eq:pwr-r-theta}) are convex. The uncertainty-induced constraints (\ref{eq:psi-jd-mat}) and (\ref{eq:psi-jr-mat}) are linear matrix inequalities with respect to the reflection coefficients. Meanwhile, the non-convexity lies in the objective function and the constraints in~(\ref{eq:pwr-j-theta}) and~(\ref{eq:phi-jd-mat}). The problem solving is analyzed as follows.

For the non-convex constraint in~(\ref{eq:pwr-j-theta}), we introduce the auxiliary variables $ \left\{  \mu_\theta, \nu_\theta, \xi_\theta,  \right\} $ such that
\begin{subnumcases}{\label{eq:mu-nu-xi}}
    \mu_{\theta}^2 \geq \left|\bm{a}^\dagger\bm{\theta} + \beta \right|, \label{eq:mu-theta}  \\
   \nu_{\theta} \leq \sqrt{\phi_{JD}},  \label{eq:nu-theta} \\
   \xi_{\theta}^2 \leq \sigma_R^2\|\bm{\theta}^\dagger\bm{E}\|^2+ \sigma_D^2 ,   \label{eq:xi-theta}
\end{subnumcases}
which allows the reformulation of~(\ref{eq:pwr-j-theta}) as
\begin{equation} \label{eq:pwr-j-theta-lb}
    P_J \geq \frac{\sqrt {P_S}}{\sqrt{c_J}} \frac{\mu_{\theta}^2}{\nu_{\theta}}- \frac{\xi_{\theta}^2}{\psi_{JD}} .  \\
\end{equation}
Then, we adopt the SCA techniques to tackle the non-convexity. Particularly, for~(\ref{eq:mu-theta}), by using the first-order lower bound of the left-hand side, we arrive at a tightened counterpart as
\begin{equation} \label{eq:mu-theta-apx}
    2\mu_{\theta}^{\circ}\mu_{\theta}-(\mu_{\theta}^{\circ})^2 \geq \left| \bm{a}^\dagger\bm{\theta} + \beta \right|,
\end{equation}
with approximation point $ \mu_{\theta}^{\circ} $. Similarly, for the non-convex constraint in~(\ref{eq:xi-theta}), we arrive an approximation as
\begin{equation} \label{eq:xi-theta-apx}
    \xi_{\theta}^2 \leq \sigma_R^2\mathsf{Tr}(\bm{E}\bm{E}^\dagger(\bm{\theta}(\bm{\theta}^{\circ})^\dagger
    +\bm{\theta}^{\circ}\bm{\theta}^\dagger-\bm{\theta}^{\circ}(\bm{\theta}^{\circ})^\dagger))+ \sigma_D^2,
\end{equation}
at the point $ \bm{\theta}^{\circ} $. While for the reformulated jamming power constraint in~(\ref{eq:pwr-j-theta-lb}), with SCA technique to tackle the non-convex part, we derive that
\begin{equation} \label{eq:pwr-j-theta-apx}
    P_J \geq \frac{\sqrt {P_S}}{\sqrt{c_J}} \frac{\mu_{\theta}^2}{\nu_{\theta}}-\frac{2\xi_{\theta}^{\circ}\xi_{\theta}\psi_{JD}^{\circ}-(\xi_{\theta}^{\circ})^2\psi_{JD}}{(\psi_{JD}^{\circ})^2}  
\end{equation}
as approximated at $ \left\{ \xi_{\theta}^{\circ}, \psi_{JD}^{\circ} \right\} $. For the non-convex constraint in~(\ref{eq:phi-jd-mat}) in a matrix inequality form, we identify the non-convexity is due to the square operation over the reflection coefficient in $ \bm{B} $. Towards this issue, we rewrite $ \bm{b}^\dagger = \left[ \bm{w}_D^\dagger \quad \bm{w}_D^\dagger\bm{H}_{RD}\bm{\Theta}  \right] = \left[ \bm{w}_D^\dagger \quad \bm{\theta}^\dagger\bm{E}  \right] $. By adopting the lower bound that $ \bm{\theta}\bm{\theta}^\dagger \succcurlyeq \bm{\theta}(\bm{\theta}^{\circ})^\dagger+\bm{\theta}^{\circ}\bm{\theta}^\dagger -\bm{\theta}^{\circ}(\bm{\theta}^{\circ})^\dagger $ approximated at $ \bm{\theta}^{\circ} $, we arrive that
\begin{equation}
\bm{b}\bm{b}^\dagger \succcurlyeq 
\underbrace{\begin{bmatrix}
    \bm{w}_D\bm{w}_D^\dagger&\bm{w}_D\bm{\theta}^\dagger\bm{E}\\
    \bm{E}^\dagger\bm{\theta}\bm{w}_D^\dagger
    &\bm{E}^\dagger(\bm{\theta}\bm{\theta}_0^\dagger
    +\bm{\theta}_0\bm{\theta}^\dagger-\bm{\theta}_0\bm{\theta}_0^\dagger)\bm{E}
\end{bmatrix}}_{\triangleq\, \bm{A}_{\theta} \left(\bm{\theta}; \bm{\theta}^{\circ}\right) }.
\end{equation}
Then, we define $ \bm{B}_{\theta} \left(\bm{\theta}; \bm{\theta}^{\circ}\right) = \bm{A}_{\theta} \left(\bm{\theta}; \bm{\theta}^{\circ}\right) \otimes\bm{I}_{N_J \times N_J} $, and rewrite the uncertainty-induced condition in~(\ref{eq:def-B}) as $ \mathrm{vec}^\dagger(\bm{H}_J) \bm{B}_{\theta} \left(\bm{\theta}; \bm{\theta}^{\circ}\right) \mathrm{vec}(\bm{H}_J)-\phi_{JD}\geq0 $. By following the similar procedure to handle the uncertainty with its bound information, we derive the condition in~(\ref{eq:B-apx-theta}), which is a linear matrix inequality with respect to $ \bm{\theta} $, as a convex replacement of the non-convex condition in~(\ref{eq:phi-jd-mat}).
\begin{figure*} 
\begin{equation}\label{eq:B-apx-theta}
\begin{aligned}
        \begin{bmatrix}
        \bm{B}_{\theta} \left(\bm{\theta}; \bm{\theta}^{\circ}\right)+\eta_1\bm{\gamma}_{JD}+\eta_2\bm{\gamma}_{JR} & \bm{B}_{\theta} \left(\bm{\theta}; \bm{\theta}^{\circ}\right)\mathrm{vec}(\hat{\bm{H}}_{J})\\
        \mathrm{vec}^\dagger(\hat{\bm{H}}_{J})\bm{B}_{\theta} \left(\bm{\theta}; \bm{\theta}^{\circ}\right)
        & \mathrm{vec}^\dagger(\hat{\bm{H}}_J)\bm{B}_{\theta} \left(\bm{\theta}; \bm{\theta}^{\circ}\right)\mathrm{vec}(\hat{\bm{H}}_J)-\phi_{JD}-\eta_1\epsilon_{JD}-\eta_2\epsilon_{JR}
    \end{bmatrix} \qquad\qquad\qquad \\ \succcurlyeq\, \bm{0}_{(N_J(N+N_D)+1)\times (N_J(N+N_D)+1)}
\end{aligned}
\end{equation}
\hrulefill
\end{figure*}
For the objective function, we also introduce the auxiliary variables $ \left\{ \phi_{\theta}, \psi_{\theta}  \right\} $ such that
\begin{subnumcases}{}
     \phi_{\theta}^2 \leq  |\bm{a}^\dagger\bm{\theta} + \beta|, \label{eq:phi-theta} \\
     \psi_{\theta} \geq  \sqrt{\psi_{JD}},  \label{eq:psi-theta}
\end{subnumcases}
followed by the lower-bounded reformulation of objective function that
\begin{equation}
    \frac{|\bm{a}^\dagger\bm{\theta} + \beta|}{\sqrt{\psi_{JD}}} \ge \frac{\phi_{\theta}^2}{\psi_{\theta}} .
\end{equation}
For the non-convexity above, we adopt necessary square operation over the inequalities and employ the SCA-based approximation as
\begin{equation} \label{eq:phi-theta-apx}
    \phi_{\theta}^4 \leq  \bm{a}^\dagger(\bm{\theta}(\bm{\theta}^{\circ})^\dagger+\bm{\theta}^{\circ}\bm{\theta}^\dagger -\bm{\theta}^{\circ}(\bm{\theta}^{\circ})^\dagger)\bm{a}
        + 2\Re\{\bm{a}^\dagger\bm{\theta}\beta\}+|\beta|^2  ,
\end{equation}
and
\begin{equation} \label{eq:psi-theta-apx}
    2\psi_{\theta}^{\circ}\psi_\theta-(\psi_{\theta}^{\circ})^2 \geq \psi_{JD},
\end{equation}
along with the lower-bounded objective function that
\begin{equation}
    \frac{\phi_{\theta}^2}{\psi_{\theta}} \geq \frac{2\phi_{\theta}^{\circ}\phi_{\theta}\psi_{\theta}^{\circ}-(\phi_{\theta}^{\circ})^2\psi_{\theta}}{(\psi_{\theta}^{\circ})^2},
\end{equation}
through the first-order approximation at the point $ \left\{ \bm{\theta}^{\circ}, \psi_{\theta}^{\circ}, \phi_{\theta}^{\circ}  \right\} $.

Consequently, for the reflection optimization in~(\ref{eq:subprob-theta}) (or equivalently in~(\ref{eq:subprob-Theta})), we derive the convexified surrogate counterpart through the reformulations as
\begin{IEEEeqnarray} {cl} \label{eq:subprob-theta-apx}
    \IEEEyesnumber \IEEEyessubnumber*
    \max_{\substack{ \bm{\theta} ,\\ \psi_{JD}, \psi_{JR}, \phi_{JD}, \\ \rho_1,\rho_2,\rho_3, \eta_1, \eta_2, \\ \mu_{\theta}, \nu_{\theta}, \xi_{\theta}, \phi_{\theta}, \psi_{\theta} } }
    \quad& \frac{2\phi_{\theta}^{\circ}\phi_{\theta}\psi_{\theta}^{\circ}-(\phi_{\theta}^{\circ})^2\psi_{\theta}}{(\psi_{\theta}^{\circ})^2}  \\
        \mathrm{s.t.}\quad&  (\ref{eq:rhos}),(\ref{eq:etas}), (\ref{eq:theta-n}), (\ref{eq:pwr-r-theta}), \nonumber \\
     & (\ref{eq:nu-theta}), (\ref{eq:mu-theta-apx}), (\ref{eq:xi-theta-apx}), (\ref{eq:pwr-j-theta-apx}), (\ref{eq:phi-theta-apx}), (\ref{eq:psi-theta-apx}) ,\nonumber \\
    & (\ref{eq:psi-jd-mat}),(\ref{eq:psi-jr-mat}), (\ref{eq:B-apx-theta}), \nonumber
\end{IEEEeqnarray}
approximated at $ \left\{ \bm{\theta}^{\circ}, \mu_{\theta}^{\circ}, \xi_{\theta}^{\circ}, \psi_{JD}^{\circ}, \psi_{\theta}^{\circ}, \phi_{\theta}^{\circ}  \right\} $. Following the idea of successive approximation, we then solve a series problem in~(\ref{eq:subprob-theta-apx}) and iteratively update the approximation point based on the obtain solution, and the solution to the original reflection problem in~(\ref{eq:subprob-theta}) can be approached upon the convergence.

\subsection{Algorithm Design}

Following the idea of BSUM framework, we can then solve the decomposed subproblems, and update the grouped optimization variables of power allocation, beamforming vectors, and active reflection matrix, and the convergence leads to the suboptimal solution to the legitimate-side anti-jamming strategy. The overall algorithm is summarized as Alg.~\ref{alg}, which adopts a Gauss-Seidel type cyclic updates. Each block is optimized while keeping the other blocks fixed, such that the interrelationship is preserved through the coefficients inherited from the current iterates.

The optimization variables (power allocation, transceiving beamforming, and active reflection) are coupled in both the SINR expression and the ARIS power constraint; hence, the three strategies are not independent decisions.
In the proposed BSUM/BSLM procedure, we adopt a Gauss--Seidel type cyclic update, where each block is optimized while keeping the other blocks fixed, such that the interrelationship is preserved through the coefficients inherited from the current iterates.
The update order can be permuted without changing feasibility and the monotonic-improvement property, provided that each block is solved exactly or via a tight lower-bounded surrogate consistent at the current point.
Nevertheless, since the overall problem is nonconvex, different update orders may lead to different stationary points and different convergence speeds; therefore, we use a fixed cyclic order for numerical stability and reproducibility.

Under the proposed BSUM decomposition, each block update solves a convex subproblem exactly (e.g., power allocation) or solves a lower-bounded convex surrogate subproblem obtained via SCA (e.g., receive beamforming and reflection). Then, the objective sequence generated by the iterations is non-decreasing and convergent. Moreover, the feasible set is compact due to the explicit power constraints and the ARIS power constraint, and hence the objective is upper bounded. Accordingly, the legitimate-side problem is monotone and bounded, and thus the convergence is guaranteed.

The complexity of the proposed algorithm is analyzed as follows. The power allocation in~(\ref{eq:subprob-pwr-equiv}) is a convex problem with fixed-number scalar variables, and the transmit beamformer update in~(\ref{eq:opt-ws}) is in closed form. Hence, the overall computational burden is dominated by the convex surrogate programs~(\ref{eq:subprob-wd-apx}) and~(\ref{eq:subprob-theta-apx}), with robust linear matrix inequality constraints. For these two optimization problems, the largest linear matrix inequality has dimension $n \triangleq N_J(N+N_D)+1$. By using an interior-point method, the per-SCA-iteration complexity is approximated as $\mathcal{O}(n^{3.5})$. Denote $I_D$ and $I_\theta$ as the numbers of inner SCA iterations for solving ~(\ref{eq:subprob-wd-apx}) and~(\ref{eq:subprob-theta-apx}), respectively. Then, the overall complexity of Alg.~1 is arrived as $\mathcal{O}\!\left(I_{\mathrm{out}}\big(I_D + I_\theta\big)\,n^{3.5}\right)$, where $ I_{\mathrm{out}} $ is the number of the required iteration for the outer BSUM loop.

Regarding the game equilibrium, denoted the obtained optimum by solving~(\ref{eq:prob-r}) as $ \left\{ P_S^{\star}, P_J^{\star}, \bm{w}_S^{\star}, \bm{w}_D^{\star}, \bm{w}_J^{\star}, \bm{\Theta}^{\star}  \right\} $, then the legitimate side will conduct robust ARIS-assisted transmissions based on this policy. Then, the jammer is aware of the legitimate transmission, and launch jamming attacks by solving~(\ref{eq:prob-j-alt}). Following the principle of Stackelberg game, the obtained jamming policy by the jammer coincide with that anticipated at the legitimate side, and therefor the equilibrium of the robust Stackelberg game is exactly $ \left\{ P_S^{\star}, P_J^{\star}, \bm{w}_S^{\star}, \bm{w}_D^{\star}, \bm{w}_J^{\star}, \bm{\Theta}^{\star}  \right\} $.

\begin{algorithm}[t] \small
\SetKwComment{Comment}{$\triangleright$\ }{}
\KwIn{Channel matrices, noise variances, maximum transmit and jamming powers, cost coefficients, ARIS power and amplification threshold;}
\KwOut{Legitimate-side strategy $\{P_S^*, \bm{w}_S^\star,\bm{w} _D^\star, \bm{\Theta}^\star\}$.}
\BlankLine
Initialize $P_S^{(0)}, \bm{w}_S^{(0)}, \bm{w}_D^{(0)},\bm{\Theta}^{(0)},  $ satisfying the constraints\;
Calculate the jammer's anticipated strategy $ \bm{w}_J^{(0)} $ and $P_J^{(0)}$ using (\ref{eq:opt-j-bf}) and~(\ref{eq:opt-j-pwr})\;
\Repeat{$|u_L^{(0)} - u_L^{\star}| < \epsilon$ }{
  Calculate initial legitimate utility $u_L^{(0)}$ \;
  \Comment{Power Allocation}
  Solve the problem in~(\ref{eq:subprob-pwr-equiv}), and get $ P_J^{\star}$ and $ \gamma^{\star} $, then obtain $ P_S^\star \gets (\gamma^\star)^2 $\;
  $ P_S^{(0)} \gets P_S^{\star} $; $ P_J^{(0)} \gets P_J^{\star} $\;
  \Comment{Transmit Beamforming Optimization}
  Given $P_S^{(0)}$, $P_J^{(0)}$, $ \bm{w}_J^{(0)} $, $ \bm{w}_D^{(0)} $,and $\bm{\Theta}^{(0)}$, update  $ \bm{w}_S $ by using~(\ref{eq:opt-ws})\;
   $ \bm{w}_S^{(0)} \gets \bm{w}_S^{\star} $\;
  \Comment{Receive Beamforming Optimization}
  \Repeat{$ \| \eta^{(0)} - \eta^\star \| \le \varepsilon $ }{
  Calculate the objective function in~(\ref{eq:subprob-wd}) as $ \eta^{(0)} $ by using $ \bm{w}_S^{(0)}$, $ \bm{w}_D^{(0)}$, and $\bm{\Theta}^{(0)}$\;
  Solve the problem in~(\ref{eq:subprob-wd-apx}) at the point $ \left\{ \bm{w}_{D}^{\circ},\mu_{D}^{\circ},  \xi_{D}^{\circ}, \psi_{JD}^{\circ}, \psi_D^{\circ}, \phi_D^{\circ} \right\} $and obtain the updated objective as $ \eta^\star $ along with the solution $\bm{w}_D^\star $, $\psi_{JD}^\star$, $\phi_{JD}^\star$,  $\rho_1^\star$, $\rho_2^\star$, $\eta_1^\star$, $\eta_2^\star$,  $\mu_D^\star$, $\nu_D^\star$, $\xi_D^\star$, $\phi_D^\star$, $\psi_D^\star$\;
  $\bm{w}_{D}^{\circ} \gets \bm{w}_D^\star$, $\mu_{D}^{\circ} \gets \mu_D^\star$,  $\xi_{D}^{\circ} \gets \xi_D^\star$, $\psi_{JD}^{\circ} \gets \psi_{JD}^\star$, $\psi_D^{\circ} \gets \psi_D^\star$, $\phi_D^{\circ} \gets \phi_D^\star$\;
  }
   $ \bm{w}_D^{(0)} \gets \bm{w}_D^{\star} $\;
  \Comment{Jamming Beamforming Determination}
   update  $ \bm{w}_J $ by using (\ref{eq:opt-j-bf}) and $\bm{w}_D^{\star} $\;
   $ \bm{w}_J^{(0)} \gets \bm{w}_J^{\star} $\;
  \Comment{Reflection Optimization}
  \Repeat{$ \| \zeta^{(0)} - \zeta^\star \| \le \varepsilon $ }{
  Calculate the objective function in~(\ref{eq:subprob-Theta}) as $ \zeta^{(0)} $ by using $ \bm{w}_S^{(0)}$, $ \bm{w}_D^{(0)}$, and $\bm{\Theta}^{(0)}$\;
  Solve the problem in~(\ref{eq:subprob-theta-apx}) at the point $ \left\{ \bm{\theta}^{\circ}, \mu_{\theta}^{\circ}, \xi_{\theta}^{\circ}, \psi_{JD}^{\circ}, \psi_{\theta}^{\circ}, \phi_{\theta}^{\circ}  \right\} $and obtain the updated objective as $ \zeta^\star $ along with the solution $\bm{\theta}^\star$, $\psi_{JD}^\star$, $\psi_{JR}^\star$, $\phi_{JD}^\star$,  $\rho_1^\star$, $\rho_2^\star$, $\rho_3^\star$, $\eta_1^\star$, $\eta_2^\star$, $\mu_{\theta}^\star$, $\nu_{\theta}^\star$, $\xi_{\theta}^\star$, $\phi_{\theta}^\star$, $\psi_{\theta}^\star$\;
  $\bm{\theta}^\star \gets \bm{\theta}^{\circ}$, $\mu_{\theta}^\star \gets \mu_{\theta}^{\circ}$, $\xi_{\theta}^\star \gets \xi_{\theta}^{\circ}$, $\psi_{JD}^\star \gets \psi_{JD}^{\circ}$, $\psi_{\theta}^\star \gets \psi_{\theta}^{\circ}$, $\phi_{\theta}^\star \gets \phi_{\theta}^{\circ}$\;
  }
  Reconstruct $ \bm{\Theta}^{\star} $ from $ \bm{\theta}^\star $, and $ \bm{\Theta}^{(0)} \gets \bm{\Theta}^\star $\;
  \Comment{Utility Update and Convergence}
  Recalculate the legitimate utility as $ u_L^\star $ by using the updated strategies\;
}
\caption{Algorithm for Legitimate-Side Strategy}
\label{alg}
\end{algorithm}

\section{Simulation Results} \label{sec:sim}

The simulation results are presented to evaluate the performance of the proposed robust jamming mitigation scheme. The following parameters are used as the default unless otherwise noted. We consider an area that the legitimate source, destination, and jammer are located at the coordinates (0, 100), (400, 0), and (100, 400), respectively (distances in meters). The ARIS is located at (200, 0) with a height of 200. The legitimate transmitter and the jammer have 4 antennas, and the legitimate destination node has 2 antennas. The ARIS has 20 reflecting elements, and the maximum amplification amplitude of each element is 10 dB, and the power limit of the ARIS is 20 dBm. The maximum jamming power and the maximum legitimate power are 10 W and 5 W, respectively. The channels on the ground suffer a path-loss exponent of 3.5 with Rayleigh fading while the channels related to the ARIS experience a path-loss exponent of 2.3 with Rician fading. The large-scale fading at a reference distance of 1 m is -20 dB and the background noise power is -140 dBW for both the receiver and the ARIS. The power cost coefficients at the legitimate source node and the jammer are 2 and 3, respectively. The channel uncertainties are defined with the uncertainty coefficient denoted by $\delta$, specified as $\epsilon_X=\delta\|\triangle\bm{H}_X\|$, $X\in\{JR,JD\}$, and the uncertainty coefficient is assumed of 0.05.% The threshold determining the convergence is 1e-3.

For performance comparison, we consider three baseline schemes. The first is the case without uncertainty, where the legitimate-side also has accurate channel information in the game. In this case, the jamming strategy remains the same, and the legitimate side solves the reduced problem in~(\ref{eq:prob-r}) while removing the uncertainty-related terms and conditions, which can be similarly solved through the BSUM framework. The second is the case of non-robust scheme, where the legitimate side takes action based on the case without uncertainty but the scenario indeed incurs uncertainty. The third remains the robust game framework but the ARIS assistance is absent.

\begin{figure}[t]
    \centering
    \includegraphics[width=0.78\columnwidth]{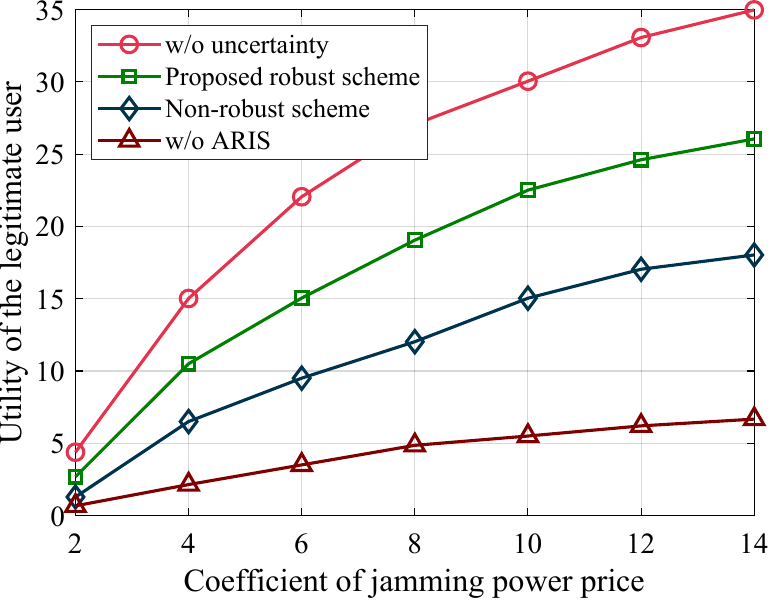}
    \caption{The utility of the legitimate user versus jamming power price.}
    \label{fig:user_utility_vs_cj}
\end{figure}

\begin{figure}[t]
    \centering
    \includegraphics[width=0.78\columnwidth]{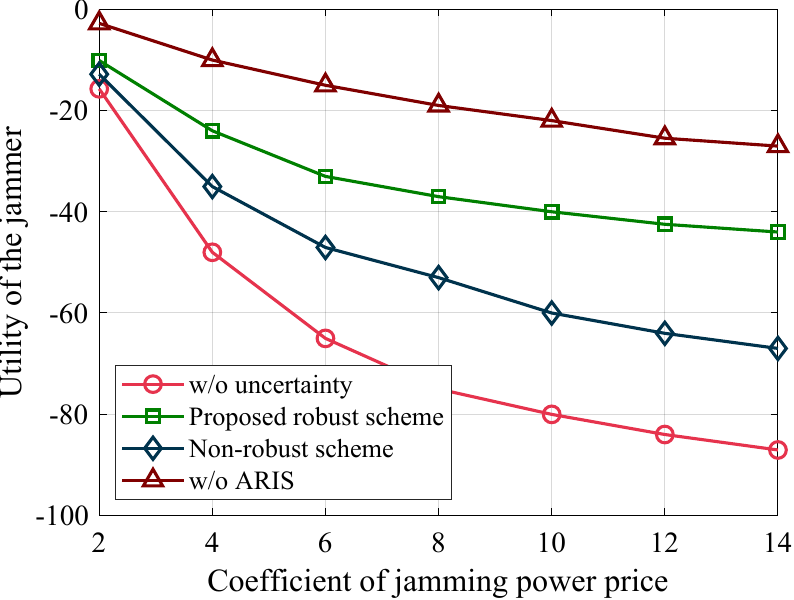}
    \caption{The utility of the jammer versus jamming power price.}
    \label{fig:jammer_utility_vs_cj}
\end{figure}

We first show the achieved utility against the jamming power price, and the results are shown in Figs.~\ref{fig:user_utility_vs_cj} and~\ref{fig:jammer_utility_vs_cj}. In Fig.~\ref{fig:user_utility_vs_cj}, the utility for the legitimate user is shown to increase for all schemes as the price becomes larger. This is as expected, since a higher jamming cost imposes a greater penalty on the jammer, induces more conservative attacks. While dominated by the case without uncertainties, the proposed scheme consistently achieves a higher utility than the other two baselines. The performance loss compared with the case of perfect information reveals the price to ensure robustness, as additional resources are devoted to ensure the worst-case protection. Meanwhile, the non-robust case has downgraded performance, because that the legitimate-side strategy targets at the case with perfect information and deviates from the actual scenario that is associated with uncertainties. For the case without ARIS, through the legitimate side still remains robust, but the wireless environment is much worsened, thus significantly degrading the performance.

For the performance of the jammer, as shown in Fig.~\ref{fig:jammer_utility_vs_cj}. Overall, it depicts an inverse trend as compared with the results in Fig.~\ref{fig:user_utility_vs_cj}, as the higher jamming price affects the jamming utility negatively. Among the considered schemes, the case without an ARIS leads to the highest jamming utility, which suggests the remarkable contribution in ARIS-assisted jamming mitigation. While for the case without uncertainty, the legitimate side has not only the advantageous position as leader in the game but also the perfect information to correctly determine the legitimate policy, and thus achieves the strongest jamming mitigation. The jamming utility under the non-robust scheme is still dominated by that under the robust scheme. This is because in this case, the legitimate-side strategy fails to match the actual scenario with uncertainties, which in return affects the jamming strategy and results in a lowered utility.

\begin{figure}[t]
    \centering
    \includegraphics[width=0.78\columnwidth]{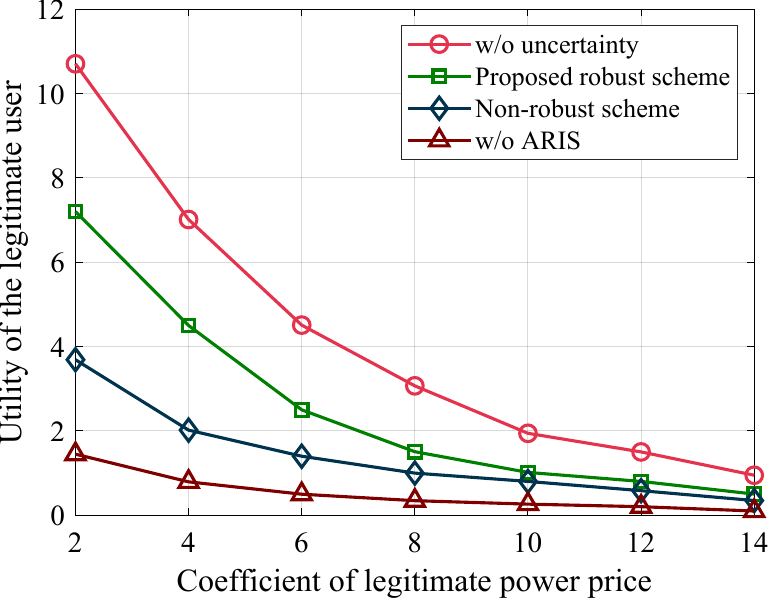}
    \caption{The utility of the legitimate user versus legitimate power price.}
    \label{fig:user_utility_vs_cs}
\end{figure}

\begin{figure}[t]
    \centering
    \includegraphics[width=0.78\columnwidth]{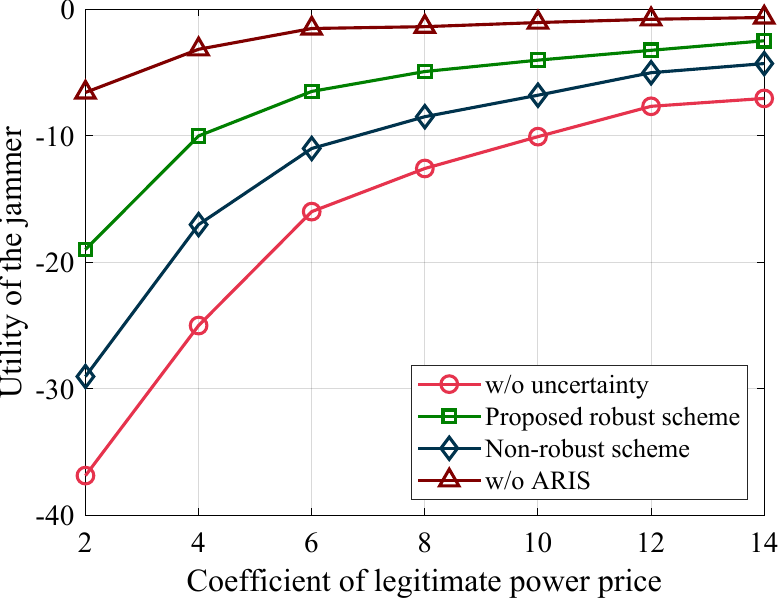}
    \caption{The utility of the jammer versus legitimate power price.}
    \label{fig:jammer_utility_vs_cs}
\end{figure}

In Figs.~\ref{fig:user_utility_vs_cs} and~\ref{fig:jammer_utility_vs_cs}, we show the achieve utility in the game with respect to the legitimate transmit power price. Overall, they show the inverse trends as compared to their counterpart results in Figs.~\ref{fig:user_utility_vs_cj} and~\ref{fig:jammer_utility_vs_cj}, since the higher legitimate power significantly confines the legitimate transmissions and thus is desired at the attacker. For the performance comparison, we can also see that the accurate information has great contribution to combat jamming attacks. While when uncertainty occurs, the proposed robust scheme ensures reliable protection of the legitimate transmissions. Also, the absence of the ARIS makes the legitimate communications significantly more vulnerable to attack. These observations underscore the efficiency of our approach, which leverages the ARIS and a robust game-based algorithm to achieve secure communication performance in the presence of attacks and uncertainties.

For the preceding four results in Figs.~\ref{fig:user_utility_vs_cj},~\ref{fig:jammer_utility_vs_cs},~\ref{fig:user_utility_vs_cs},~\ref{fig:jammer_utility_vs_cs}, we have an interesting observation, that the proposed robust scheme yields a higher utility for not only the legitimate user but also for the jammer when compared to the non-robust scheme, regardless of the power price at both sides. The seemingly counter-intuitive results are fully reasonable within the formulated Stackelberg game framework. The non-robust design induces a legitimate strategy based on flawed channel estimates, committing to a transmission plan that is mismatched to the true channel environment. The jammer, acting as a rational follower with perfect channel knowledge, observes this misconfigured strategy and is likely to be provoked into an aggressively high-power attack to exploit the legitimate-side vulnerability. This high jamming power results in a severe cost penalty for the jammer and thus lowers its own utility.
In a nutshell, the mismatched legitimate strategy also induces a mismatched jamming policy, following the sequential decision-making process in the Stackelberg game, and further downgrades the achievable utility at both game players.

\begin{figure}[t]
    \centering
    \includegraphics[width=0.78\columnwidth]{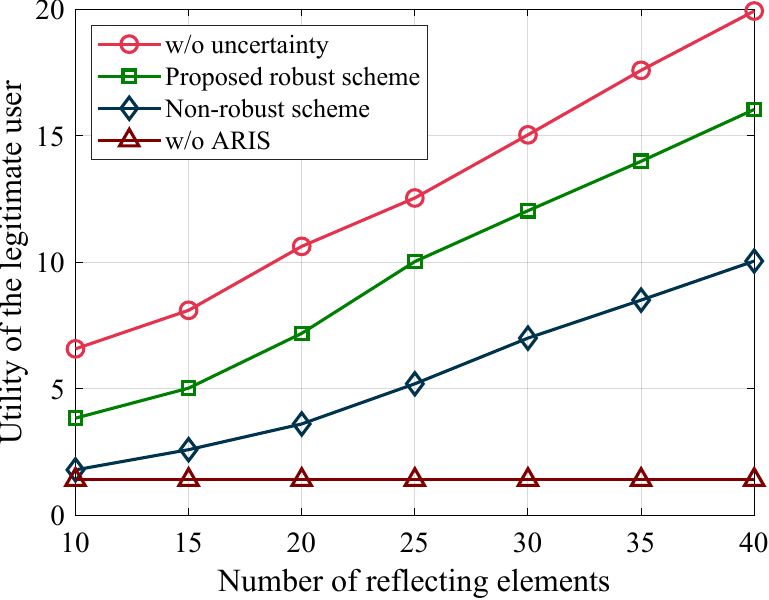}
    \caption{The utility of the legitimate user versus number of reflecting elements.}
    \label{fig:user_utility_vs_n}
\end{figure}

\begin{figure}[t]
    \centering
    \includegraphics[width=0.78\columnwidth]{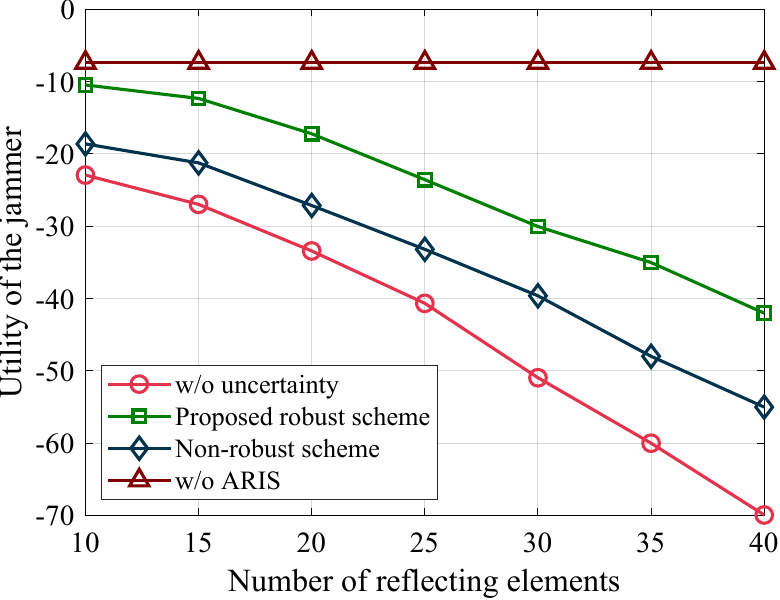}
    \caption{The utility of the jammer versus number of reflecting elements.}
    \label{fig:jammer_utility_vs_n}
\end{figure}

In Figs.~\ref{fig:user_utility_vs_n} and~\ref{fig:jammer_utility_vs_n}, we show the achieved utility of the game players with respect to the number of reflecting elements at the ARIS. In Fig.~\ref{fig:user_utility_vs_n}, the legitimate-side utility of all schemes employing an ARIS increases monotonically with an increasing number of reflecting elements. This is because a larger surface provides a higher degree of freedom for environment programming, leading to a superior beamforming gain that can be harnessed to enhance the legitimate signal while alleviating the malicious jamming. More importantly, the signal amplification at the ARIS can even strengthen such a desired effect.
In contrast, the utility of the case without ARIS remains flat, as its performance is independent of the reflection. In this regard, the significant performance gap between the cases with and without ARIS clearly demonstrate the critical role of ARIS in anti-jamming communications, for which with more reflecting elements, i.e., higher capability of the ARIS, this contribution becomes more prominent. Moreover, the proposed robust scheme effectively leverages the reflection to improve its utility, consistently outperforming the non-robust baseline approach and closing the gap to the case with perfect information as the ARIS grows larger.

Fig.~~\ref{fig:jammer_utility_vs_n} presents the corresponding utility for the jammer as a function of the number of ARIS elements. Overall, the figure largely shows the inverse trend as compared with Fig.~\ref{fig:user_utility_vs_n}, that for all ARIS-assisted schemes, the jammer's utility decreases as the ARIS becomes larger. A larger surface allows to more effectively nullify the incoming jamming signals and thereby diminish the jammer's impact. As compared with our proposed scheme, the improved channel information accuracy (c.f., the case without uncertainty) and matched response to the environment (c.f., the non-robust) further improve the performance gain delivered by the ARIS. This result is crucial as it demonstrates that our approach not only improves the legitimate-side performance but also makes the environment more hostile and challenging for the adversary.

\begin{figure}[t]
    \centering
    \includegraphics[width=0.78\columnwidth]{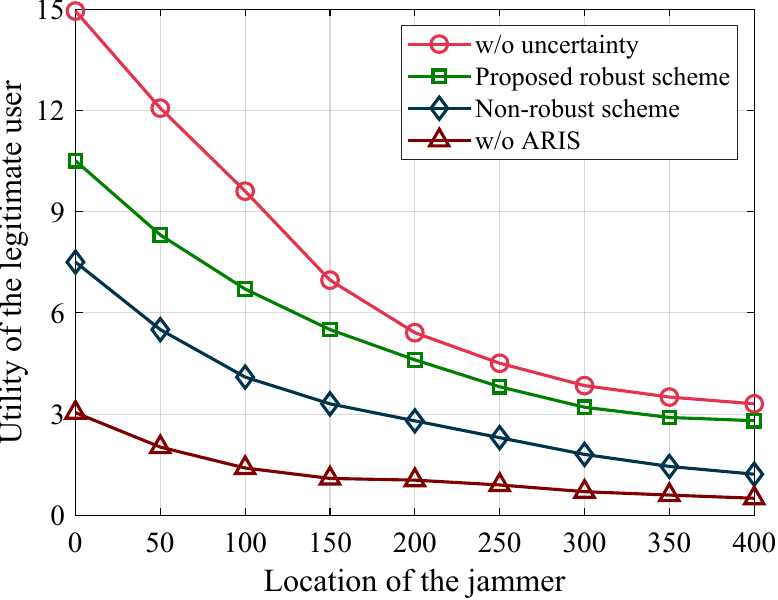}
    \caption{The utility of the legitimate user versus location of the jammer.}
    \label{fig:user_utility_vs_loc}
\end{figure}

\begin{figure}[t]
    \centering
    \includegraphics[width=0.78\columnwidth]{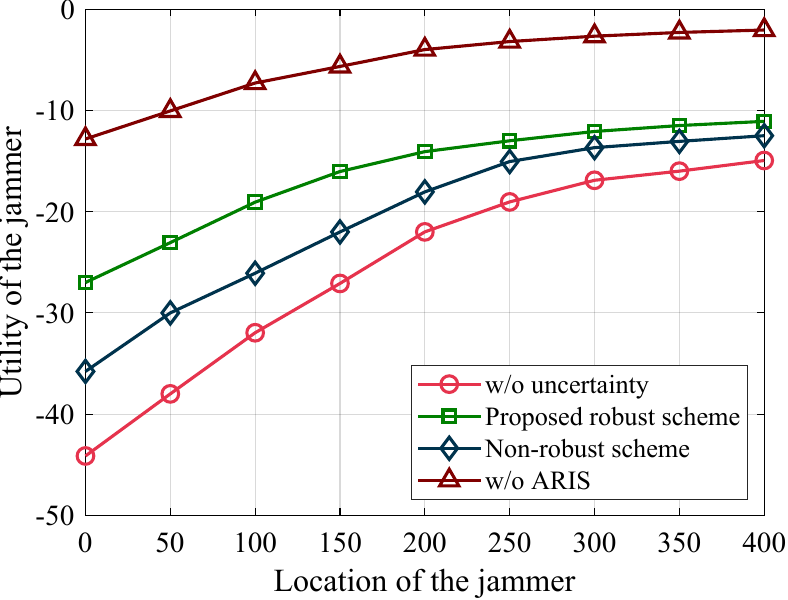}
    \caption{The utility of the jammer versus its location.}
    \label{fig:jammer_utility_vs_loc}
\end{figure}

The anti-jamming performance considering the location of the jammer is shown in Figs.~\ref{fig:user_utility_vs_loc} and~\ref{fig:jammer_utility_vs_loc}, for which as the jammer moves from the left to the right, it deviates farther from the legitimate source and locates closer to the destination. As seen in Fig.~\ref{fig:user_utility_vs_loc}, the utility of the legitimate user deteriorates as the jammer moves to a more advantageous attack position. Despite this challenging scenario, our proposed robust scheme consistently maintains a significantly higher utility compared to both the baselines of non-robust scheme and without an ARIS across all tested locations. This demonstrates the resilience of our approach in defending against a spatially optimized adversary. Also, we notice that facing a more challenging attacker, the advantage due to perfect information is compromised as the gap between the case without uncertainty and our proposal becomes narrowed. In contrast, the performance superiority of our approach as compared with the non-robust case remains evident, implying the strong resilience of the proposed method to defend against jamming in uncertain jamming environment.

Fig.~\ref{fig:jammer_utility_vs_loc} shows the utility of the jammer as a function of its location. As a mirror to the previous results, the jammer's utility increases as it moves closer to the destination, as its transmissions become more effective at disrupting the legitimate reception. Nevertheless, the proposed robust scheme implements effective defense and consistently suppresses the jammer's utility, regardless of the jammer's position. In contrast, the case without an ARIS allows a highly effective jamming attacks. This particularly reveals that the proposed approach with a combination of the ARIS and robust design provides a comprehensive security solution that is not easily defeated by the spatial movement of the adversary.

\section{Conclusion} \label{sec:con}

In this paper, we have developed a robust anti-jamming framework for an ARIS-assisted system facing an adaptive jammer in the presence of channel uncertainties, by modeling the strategic conflict as a robust Stackelberg game. Following the principle of backward induction, we have proposed an iterative algorithm to find the optimal strategies for the legitimate side as well as the jammer. The numerical results have validated the superiority of the proposed approach, demonstrating that it has effectively mitigated the jamming attacks under varying power prices, reflecting sizes, and attacking locations, and has reliably combated the uncertainties to achieve robust protection of the legitimate transmissions.

\bibliographystyle{IEEEtran}
\bibliography{main}

@ARTICLE{back-jam,
  author={Pirayesh, Hossein and Zeng, Huacheng},
  journal={IEEE Commun. Surveys Tuts.}, 
  title={Jamming Attacks and Anti-Jamming Strategies in Wireless Networks: A Comprehensive Survey}, 
  year={2022},
  volume={24},
  number={2},
  pages={767-809},
  doi={10.1109/COMST.2022.3159185}}

@ARTICLE{back-anti-jam,
  author={Han, Chen and An, Kang and Lin, Zhi and Chatzinotas, Symeon and Wang, Jiangzhou},
  journal={IEEE Netw.}, 
  title={Endogenous Anti-Jamming Communications for {SAGIN}: A Network Perspective}, 
  year={2025},
  
  note={to appear},
  doi={10.1109/MNET.2025.3551248}}

@ARTICLE{back-adp-jam,
  author={Zhang, Pinchang and Dong, Jiankuo and He, Ji and Liu, Jun and Xiao, Fu},
  journal={IEEE Trans. Veh. Technol.}, 
  title={Adaptive Jamming Attack Detection Under Noise Uncertainty in mmWave Massive {MIMO} Systems}, 
  year={2023},
  volume={72},
  number={7},
  pages={9002-9016},
  doi={10.1109/TVT.2023.3247488}}

@ARTICLE{back-jam-vis,
  author={Li, Guyue and Staat, Paul and Li, Haoyu and Heinrichs, Markus and Zenger, Christian and Kronberger, Rainer and Elders-Boll, Harald and Paar, Christof and Hu, Aiqun},
  journal={IEEE Trans. Inf. Forensics Sec.}, 
  title={{RIS}-Jamming: Breaking Key Consistency in Channel Reciprocity-Based Key Generation}, 
  year={2024},
  volume={19},
  number={},
  pages={5090-5105},
  doi={10.1109/TIFS.2024.3389569}}

@ARTICLE{jam-survey,
  author={Aygur, Michael and Kandeepan, Sithamparanathan and Giorgetti, Andrea and Al-Hourani, Akram and Arbon, Edward and Bowyer, Mark},
  journal={IEEE Commun. Surveys Tuts.}, 
  title={Narrowband Interference Mitigation Techniques: A Survey}, 
  year={2025},

  note={to appear},
  doi={10.1109/COMST.2025.3531428}}

@ARTICLE{jam-spec,
  author={Li, Wen and Chen, Jin and Liu, Xin and Wang, Ximing and Li, Yangyang and Liu, Dianxiong and Xu, Yuhua},
  journal={IEEE Wireless Commun.}, 
  title={Intelligent Dynamic Spectrum Anti-Jamming Communications: A Deep Reinforcement Learning Perspective}, 
  year={2022},
  volume={29},
  number={5},
  pages={60-67},
  doi={10.1109/MWC.103.2100365}}

@ARTICLE{jam-sec,
  author={Dong, Limeng and Huo, Yiran and Yan, Wanyu and Tang, Xiao and Li, Yong and Cheng, Wei},
  journal={IEEE Trans. Cog. Commun. Netw.}, 
  title={Joint Secure Transmission Enhancement of Primary and Secondary Users in {RIS} Aided Spectrum Sharing Cognitive Radio Networks}, 
  year={2025},
  volume={11},
  number={6},
  pages={3744-3760},
  doi={10.1109/TCCN.2025.3541219}}

@ARTICLE{jam-aris,
  author={Tang, Xiao and Ma, Zhen and Li, Bin and Li, Cong and Du, Qinghe and Niyato, Dusit and Han, Zhu},
  journal={IEEE Trans. Commun.}, 
  title={Active {RIS}-Aided Anti-Jamming Wireless Communications: A Stackelberg Game Perspective}, 
  year={2026},
  volume={74},
  number={},
  pages={2612-2625},
  doi={10.1109/TCOMM.2025.3646817}}

@ARTICLE{game-jam,
  author={Jia, Luliang and Qi, Nan and Chu, Feihuang and Fang, Shengliang and Wang, Ximing and Ma, Shuli and Feng, Shuo},
  journal={IEEE Commun. Mag.}, 
  title={Game-Theoretic Learning Anti-Jamming Approaches in Wireless Networks}, 
  year={2022},
  volume={60},
  number={5},
  pages={60-66},
  doi={10.1109/MCOM.001.00496}}

@ARTICLE{game-reac,
  author={Van Huynh, Nguyen and Nguyen, Diep N. and Hoang, Dinh Thai and Vu, Thang X. and Dutkiewicz, Eryk and Chatzinotas, Symeon},
  journal={IEEE Trans. Wireless Commun.}, 
  title={Defeating Super-Reactive Jammers With Deception Strategy: Modeling, Signal Detection, and Performance Analysis}, 
  year={2022},
  volume={21},
  number={9},
  pages={7374-7390},
  doi={10.1109/TWC.2022.3158189}}

@ARTICLE{game-gam,
  author={Xu, Yifan and Xu, Yuhua and Ren, Guochun and Chen, Jin and Yao, Changhua and Jia, Luliang and Liu, Dianxiong and Wang, Ximing},
  journal={IEEE Trans. Inf. Forensics Sec.}, 
  title={Play it by Ear: Context-Aware Distributed Coordinated Anti-Jamming Channel Access}, 
  year={2021},
  volume={16},
  number={},
  pages={5279-5293},
  doi={10.1109/TIFS.2021.3128249}}

@ARTICLE{game-ris,
  author={Xu, Youyun and Wang, Weinan and Li, Tianyou},
  journal={IEEE Trans. Green Commun. Netw.}, 
  title={A Communication Confrontation Based on the {Stackelberg} Game: {IRS} Aids for Improving Anti-Jamming Capability}, 
  year={2024},
  note={to appear},
  doi={10.1109/TGCN.2024.3515637}}

@ARTICLE{rob-back,
  author={Liu, Yiyuan and Wang, Jinlong and Zhang, Xiaokai and Li, Guoxin and Xu, Yuhua},
  journal={IEEE Trans. Veh. Technol.}, 
  title={Spatial Anti-Jamming Based on Low Complexity Robust Beamforming Via Orthogonal Projection}, 
  year={2025},

  note={to appear},
  doi={10.1109/TVT.2025.3553109}}

@ARTICLE{rob-ris,
  author={Ye, Runlong and Peng, Yuyang and Al-Hazemi, Fawaz and Boutaba, Raouf},
  journal={IEEE Trans. Commun.}, 
  title={A Robust Cooperative Jamming Scheme for Secure {UAV} Communication via Intelligent Reflecting Surface}, 
  year={2024},
  volume={72},
  number={2},
  pages={1005-1019},
  doi={10.1109/TCOMM.2023.3325488}}

@ARTICLE{rob-rob,
  author={Qi, Xiaolei and Peng, Mugen and Zhang, Hongming and Kong, Xianghao},
  journal={IEEE Trans. Wireless Commun.}, 
  title={Anti-Jamming Hybrid Beamforming Design for Millimeter-Wave Massive {MIMO} Systems}, 
  year={2024},
  volume={23},
  number={8},
  pages={9160-9172},
  doi={10.1109/TWC.2024.3359286}}

@ARTICLE{rob-todo,
  author={Sun, Yifu and An, Kang and Yu, Miao and Hu, Yihua and Zhu, Yonggang and Lin, Zhi and Xiao, Ming and Al-Dhahir, Naofal and Niyato, Dusit and Wang, Jiangzhou},
  journal={IEEE J. Sel. Areas Commun.}, 
  title={Dual-Polarized Stacked Metasurface Transceiver Design With Rate Splitting for Next-Generation Wireless Networks}, 
  year={2025},
  volume={43},
  number={3},
  pages={811-833},
  doi={10.1109/JSAC.2025.3531526}}

@ARTICLE{ris-back-jam1,
  author={Ma, Jianhui and Li, Qiang and Liu, Zilong and Du, Linsong and Chen, Hongyang and Ansari, Nirwan},
  journal={IEEE Trans. Wireless Commun.}, 
  title={Jamming Modulation: An Active Anti-Jamming Scheme}, 
  year={2023},
  volume={22},
  number={4},
  pages={2730-2743},
  doi={10.1109/TWC.2022.3213572},
  ISSN={1558-2248},
  month={April},}

@ARTICLE{ris-fair,
  author={Liu, Jun and Yang, Gang and Liang, Ying-Chang and Yuen, Chau},
  journal={IEEE Trans. Commun.}, 
  title={Max-Min Fairness in {RIS}-Assisted Anti-Jamming Communications: Optimization Versus Deep Reinforcement Learning Approaches}, 
  year={2024},
  volume={72},
  number={7},
  pages={4476-4492},
  doi={10.1109/TCOMM.2024.3371359}}

@ARTICLE{ris-wpt,
  author={Chu, Zheng and Chieng, David and Foong Kwong, Chiew and Jin, Huan and Zhu, Zhengyu and Huang, Chongwen and Yuen, Chau},
  journal={IEEE Trans. Inf. Forensics Sec.}, 
  title={Throughput Improvement for {RIS}-Empowered Wireless Powered Anti-Jamming Communication Networks ({WPAJCN})}, 
  year={2025},
  volume={20},
  number={},
  pages={4622-4637},
  doi={10.1109/TIFS.2025.3563818}}

@ARTICLE{ris-isac,
  author={Xu, Jinlei and Li, Dongdong and Zhu, Zhengyu and Yang, Zhutian and Zhao, Nan and Niyato, Dusit},
  journal={IEEE Trans. Commun.}, 
  title={Anti-Jamming Design for Integrated Sensing and Communication via Aerial {IRS}}, 
  year={2024},
  volume={72},
  number={8},
  pages={4607-4619},
  doi={10.1109/TCOMM.2024.3375809}}

@ARTICLE{ris-ln,
  author={Tang, Xiao and Zhao, Kexin and Shen, Chao and Lin, Chenhao and Liu, Shuai and Wang, Bohui and Niyato, Dusit and Han, Zhu},
  journal={IEEE Trans. Wireless Commun.}, 
  title={Graph Attention Network-Driven Hierarchical Learning for Anti-Jamming {UAV} Communications}, 
  year={2026},
  volume={25},
  number={},
  pages={5432-5445},
  doi={10.1109/TWC.2025.3618614}}

@ARTICLE{ris-opt,
  author={Yang, Yan and Dang, Shuping and Wen, Miaowen and Ai, Bo and Hu, Rose Qingyang},
  journal={IEEE Trans. Wireless Commun.}, 
  title={Blockage-Aware Robust Beamforming in {RIS}-Aided Mobile Millimeter Wave {MIMO} Systems}, 
  year={2024},
  volume={23},
  number={11},
  pages={16906-16921},
  doi={10.1109/TWC.2024.3447895}}

@article{ris-cas,
title = {{UAV}-enabled aerial active {RIS} with learning deployment for secured wireless communications},
journal = {Chin. J. Aeronaut.},
volume={38},
pages={103383},
year = {2025},
issn = {1000-9361},
doi = {https://doi.org/10.1016/j.cja.2024.103383},
author = {Xiao Tang and Zhihao Xiong and Limeng Dong and Ruonan Zhang and Qinghe Du},
}

@ARTICLE{ris-a,
  author={Dong, Limeng and Li, Yong and Cheng, Wei and Huo, Yiran},
  journal={IEEE Trans. Veh. Technol.}, 
  title={Robust and Secure Transmission Over Active Reconfigurable Intelligent Surface Aided Multi-User System}, 
  year={2023},
  volume={72},
  number={9},
  pages={11515-11531},
  doi={10.1109/TVT.2023.3262959}}

@ARTICLE{ris-sus,
  author={Cao, Yang and Cheng, Wenchi and Wang, Jingqing and Zhang, Wei},
  journal={IEEE Systems J.}, 
  title={Self-Sustainable Active Reconfigurable Intelligent Surfaces for Antijamming in Wireless Communications}, 
  year={2024},
  volume={18},
  number={4},
  pages={2133-2144},
  doi={10.1109/JSYST.2024.3470133}}

@ARTICLE{ga-zero1,
  author={Shi, Bai and Shao, Huaizong and Lin, Jingran and Zhao, Shenglan and Wang, Shafei},
  journal={IEEE Trans. Inf. Forensics Sec.}, 
  title={Jamming the Relay-Assisted Multi-User Wireless Communication System: A Zero-Sum Game Approach}, 
  year={2023},
  volume={18},
  number={},
  pages={3198-3211},
  doi={10.1109/TIFS.2023.3277222}}

@ARTICLE{ga-sta1,
  author={Lin, Zhiping and Xiao, Liang and Chen, Hongyi and Lv, Zefang and Zhu, Yunjun and Zhang, Yanyong and Liu, Yong-Jin},
  journal={IEEE Trans. Wireless Commun.}, 
  title={Edge-Assisted Collaborative Perception Against Jamming and Interference in Vehicular Networks}, 
  year={2025},
  volume={24},
  number={1},
  pages={860-874},
  doi={10.1109/TWC.2024.3510601}}

@ARTICLE{ga-sta2,
  author={Liu, Jianhua and Wang, Xin and Shen, Shigen and Fang, Zhaoxi and Yu, Shui and Yue, Guangxue and Li, Minglu},
  journal={IEEE Internet Things J.}, 
  title={Intelligent Jamming Defense Using {DNN} {Stackelberg} Game in Sensor Edge Cloud}, 
  year={2022},
  volume={9},
  number={6},
  pages={4356-4370},
  doi={10.1109/JIOT.2021.3103196}}

@ARTICLE{ga-ris1,
  author={Hou, Zhifeng and Huang, Yuzhen and Chen, Jin and Li, Guoxin and Guan, Xinrong and Xu, Yifan and Chen, Runfeng and Xu, Yuhua},
  journal={IEEE Internet Things J.}, 
  title={Joint {IRS} Selection and Passive Beamforming in Multiple {IRS}-{UAV}-Enhanced Anti-Jamming {D2D} Communication Networks}, 
  year={2023},
  volume={10},
  number={22},
  pages={19558-19569},
  doi={10.1109/JIOT.2023.3281608}}

@ARTICLE{ga-ris2,
  author={Li, Baogang and Shi, Tai and Zhao, Wei and Wang, Ning},
  journal={IEEE Trans. Commun.}, 
  title={Reinforcement Learning-Based Intelligent Reflecting Surface Assisted Communications Against Smart Attackers}, 
  year={2022},
  volume={70},
  number={7},
  pages={4771-4779},
  doi={10.1109/TCOMM.2022.3178755}}

@ARTICLE{ga-rob1,
  author={Shen, Zhexian and Xu, Kui and Xia, Xiaochen},
  journal={IEEE Trans. Inf. Forensics Sec.}, 
  title={Beam-Domain Anti-Jamming Transmission for Downlink Massive {MIMO} Systems: A {Stackelberg} Game Perspective}, 
  year={2021},
  volume={16},
  number={},
  pages={2727-2742},
  doi={10.1109/TIFS.2021.3063632}}

@ARTICLE{ga-rob2,
  author={Han, Hao and Xu, Yuhua and Li, Wen and Wang, Ximing and Xu, Yifan and Zhang, Xiaokai and Gao, Yong},
  journal={IEEE Internet Things J.}, 
  title={Robust Spectrum Access Scheme Against Diverse Jamming Policies: A Prioritized Fictitious Rival-Play-Based Approach}, 
  year={2025},
  volume={12},
  number={1},
  pages={1-17},
  doi={10.1109/JIOT.2024.3459935}}

@ARTICLE{ga-rob-ris2,
  author={Sun, Yifu and Zhu, Yonggang and An, Kang and Zheng, Gan and Chatzinotas, Symeon and Wong, Kai-Kit and Liu, Pengtao},
  journal={IEEE Trans. Veh. Technol.}, 
  title={Robust Design for {RIS}-Assisted Anti-Jamming Communications With Imperfect Angular Information: A Game-Theoretic Perspective}, 
  year={2022},
  volume={71},
  number={7},
  pages={7967-7972},
  doi={10.1109/TVT.2022.3166656}}

@ARTICLE{ga-rob-ris1,
  author={Zou, Chao and An, Kang and Lin, Zhi and He, Yuanzhi and Zhong, Xudong and Zheng, Gan and Al-Dhahir, Naofal},
  journal={IEEE Commun. Lett.}, 
  title={Multi-Layer {RIS}-Assisted Anti-Jamming Communications: A Hierarchical Game Learning Approach}, 
  year={2023},
  volume={27},
  number={11},
  pages={2998-3002},
  doi={10.1109/LCOMM.2023.3321897}}

@article{E,
author = {R. Lucchetti and F. Mignanego and G. Pieri },
title = {Existence theorems of equilibrium points in {Stackelberg}},
journal = {Optimization},
volume = {18},
number = {6},
pages = {857-866},
year = {1987},
}

@book{HanGameBook,
author = "Zhu Han and Dusit Niyato and Walid Saad and Tamer Ba{\c{s}}ar",
title = {Game Theory for Next-Generation Wireless and Communication Networks: Modeling, Analysis, and Design}, 
publisher = {Cambridge University Press}, 
address = {Cambridge, UK},
year = {2019}
}

@ARTICLE{math1,
  author={Gharavol, Ebrahim A. and Larsson, Erik G.},
  journal={IEEE Trans. Signal Process.}, 
  title={The Sign-Definiteness Lemma and Its Applications to Robust Transceiver Optimization for Multiuser {MIMO} Systems}, 
  year={2013},
  volume={61},
  number={2},
  pages={238-252},
  doi={10.1109/TSP.2012.2222379}}

@article{math2,
  title={Strong duality in nonconvex quadratic optimization with two quadratic constraints},
  author={Beck, Amir and Eldar, Yonina C},
  journal={SIAM J. Optim.},
  volume={17},
  number={3},
  pages={844--860},
  year={2006},
  publisher={SIAM}
}

@ARTICLE{ris-sem,
  author={Sun, Yifu and Lin, Zhi and An, Kang and Li, Dong and Li, Cheng and Zhu, Yonggang and Wing Kwan Ng, Derrick and Al-Dhahir, Naofal and Wang, Jiangzhou},
  journal={IEEE J. Sel. Areas Commun.}, 
  title={Multi-Functional {RIS}-Assisted Semantic Anti-Jamming Communication and Computing in Integrated Aerial-Ground Networks}, 
  year={2024},
  volume={42},
  number={12},
  pages={3597-3617},
  doi={10.1109/JSAC.2024.3459028}}

@ARTICLE{s-suv,
  author={Sun, Yifu and An, Kang and Zhu, Yonggang and Guo, Kefeng and Lin, Zhi and Nauman, Ali and Jamshed, Muhammad Ali},
  journal={IEEE Netw.}, 
  title={Unlocking Potentials of {RIS} for Endogenous Anti-Jamming Communications in Internet of Energy: From Passive Defense to Active Immunity}, 
  year={2025},
  note={to appear},
  doi={10.1109/MNET.2025.3629968}}

\end{document}